%% file: AA2.tex
\documentclass[letter,referee,draft]{aa}

\usepackage{ifdraft} \usepackage{ifpdf} \usepackage[latin1]{inputenc}
\usepackage[english]{babel}

\usepackage{natbib}
\bibpunct{(}{)}{;}{a}{}{,} 

\ifpdf
    \usepackage[pdftex,final]{graphicx}
    \DeclareGraphicsExtensions{.pdf,.jpeg,.png}
\else
    \usepackage[dvips,final]{graphicx}
    \usepackage{psfrag}
    \DeclareGraphicsExtensions{.eps}
\fi

\usepackage{times}
\usepackage{color}
\usepackage{subfigure}
\usepackage{multirow}
\usepackage{amsmath}
\usepackage{dsfont}
\usepackage{algpseudocode}
\usepackage[autolanguage]{numprint}
\usepackage[draft]{fixme} 
\usepackage{soul} 
\usepackage{tikz}\usetikzlibrary{shapes,snakes}

\input alphabet.tex 
\input abrege.tex 
\input abrmath.tex 

\input notations
\def\Ciel{\Xc}

\def\eR{\mathds{R}}

\def\prior{prior\XS}
\def\post{posterior\XS}
\def\Post{Posterior\XS}

\begin{document}

\title{Estimating hyperparameters and instrument parameters in
  regularized inversion. \\ Illustration for SPIRE/Herschel map
  making.}

\titlerunning{Unsupervised and myopic inverse problems.}

\author{F. Orieux\inst{1,3} \and J.-F. Giovannelli\inst{2,3} \and
  T. Rodet\inst{3} \and A. Abergel\inst{4}\\ }

\institute{
  Institut d'Astrophysique de Paris (\cnrs~--~Univ. Paris~6), 75\,014
  Paris, France. \texttt{orieux@iap.fr}
  \and
  Univ. Bordeaux, IMS, UMR~5218, F-33\,400 Talence,
  France. \email{Giova@IMS-Bor\-deaux.fr}
  \and
  Laboratoire des Signaux et Syst\`emes
  (\cnrs~--~Sup\'elec~--~Univ. Paris-Sud 11), 91\,192 Gif-sur-Yvette,
  France. \email{orieux,rodet@lss.supelec.fr}
  \and
  Institut d'Astrophysique Spatiale (\cnrs~--~Univ. Paris-Sud 11),
  91\,405 Orsay, France. \email{abergel@ias.u-psud.fr}
}

\abstract{We describe regularized methods for image reconstruction and focus on the question of hyper\-parameter and instrument parameter estimation, i.e. unsupervised and myopic problems. We developed a Bayesian framework that is based on the \post density for all unknown quantities, given the observations. This density is explored by a Markov Chain Monte-Carlo sampling technique based on a Gibbs loop and including a Metropolis-Hastings step. The numerical evaluation relies on the SPIRE instrument of the Herschel observatory. Using simulated and real observations, we show that the hyperparameters and instrument parameters are correctly estimated, which opens up many perspectives for imaging in astrophysics.

  \keywords{Techniques: inverse problem, Bayesian regularization,
    hyperparameter estimation, instrument parameter estimation,
    semi-blind, myopic, autocalibration, image processing,
    deconvolution, super-resolution.}

}

\maketitle

\section{Unsupervised myopic inversion}

The agreement of physical models and observations is a crucial
question in astrophysics, however, observation instruments inevitably
have defects and limitations (limited pass-band, non-zero response
time, attenuation, error and uncertainty, etc.). Their inversion by
numerical processing must, as far as possible, be based on an
instrument model that includes a description of these defects and
limitations. The difficulties of such \emph{inverse problems}, and
notably their often \emph{ill-posedness}, were well identified several
decades ago in various communities: \emph{signal and image processing}
and \emph{statistics}, and also \emph{mathematical physics} and
\emph{astrophysics}. It seems pertinent to take advantage of the
knowledge amassed by these communities concerning both the analysis of
the problems and their solutions.

The \emph{ill-posedness} comes from a deficit of available information
(and not only from a ``simple numerical problem''), which becomes all
the more marked as resolution requirements increase. The inversion
methods must therefore take other information into account to
compensate for the deficits in the observations: this is known as
\emph{regularization}. Each reconstruction method is thus specialised
for a certain class of objects (point sources, diffuse emission,
superposition of the two, etc.) according to the information accounted
for. Consequently, in as much as it relies on various sources of
information, each method is based on a trade-off, which usually
requires the setting of \emph{hyperparameters}, denoted by $\xib$ in
the following. The question of their automatic tuning, namely
\emph{unsupervised inversion}, has been extensively studied and
numerous attempts investigate statistical approaches: approximated,
pseudo or marginal likelihood, in a Bayesian or non-Bayesian sense,
EM, SEM and SAEM algorithms, etc. The reader may consult papers such
as~\citep{Zhou97,Figueiredo97,Saquib98,Descombes99a,Molina99,
  Lanterman00,Pascazio03,Blanc03,Chantas07,Giovannelli08,Babacan10,Orieux10}
and reference books such as~\cite[Part.VI]{Winkler03},
\cite[Ch.7]{Li01} or~\cite[Ch.8]{Idier08}. Alternative methods are
based on the L-curve~\citep{Hansen92,Wiegelmann03} or on generalised
cross-validation~\citep{Golub79,Fortier93,Pichon06}.

The construction of maps of high resolution and accuracy relies on
increasingly complex instruments. So, inversion methods require
instrument models that faithfully reflect the physical reality to
distinguish, in the observations, between what is caused by the
instrument and what is due to the actual sky. Then, a second set of
parameters comes into play: the \emph{instrument parameters}, denoted
by $\etab$ in the following, such as lobe width, amplitude of
secondary lobes, response time, or gain. Their values are of prime
importance and their settings are generally based on dedicated
observation and rely on models and/or calibrations that inevitably
contain errors. For example, the lobe widths are usually determined
from a specific observation in a spectral band of non-zero width;
consequently the result depends on the source spectrum. Correction
factors can be applied but, naturally, they also contain errors when
the source spectrum is poorly known or unknown. In contrast, our aim
is to achieve \emph{myopic inversion}, i.e. to estimate the instrument
parameters without dedicated observation. The question arises in
various fields: optical imaging~\citep{Pankajakshan09}, interferometry
\citep{Thiebaut08}, satellite observation~\citep{Jalobeanu02a},
magnetic resonance force microscopy~\citep{Dobigeon09}, fluorescence
microscopy~\citep{Zhang06}, deconvolution~\citep{Orieux10a}, etc. A
similar problem deals with non-parametric intrument response
(\emph{blind inversion}), for which the literature is also very abundant:
\citep{Mugnier04,Thiebaut95,Fusco99,Conan98a} in astronomy and
\citep{Lam00,Likas04,Molina06,Bishop08,Xu09} in the signal-image
literature represent examples. The present paper is devoted to
parameter estimation for the instrument model developed in our
previous paper~\citep{Orieux12}, based on an accurate instrument
model.

A threefold problem has to be solved: from a unique observation,
estimate the hyperparameters the instrument parameters and the
map. This is referred to as \emph{unsupervised and myopic
  inversion}. From the methodological point of view, the proposed
inversion method comes within a Bayesian approach~\citep{Idier08}. In
this family, we find the classic Wiener and Kalman methods that
calculate the expectation or the maximizer of a \post density. In an
equivalent way, the Phillips-Twomey-Tikhonov methods calculate the
minimizer of a least-squares criterion with quadratic
penalization. These methods are based on a second-order analysis
(Gaussian models, quadratic criteria) and lead to linear
processing. The work proposed here is in a similar methodological vein
as far as estimating the map goes; however, the contribution concerns
the estimation of the hyperparameters and instrument parameters. We
resort to an entirely Bayesian approach (also called
\textit{full-Bayes}) that models the information for each variable
(observations, unknown map as well as hyperparameters and instrument
parameters) through a probability density. Based on an \apost
distribution for all the unknown variables, the proposed method
jointly estimates the instrument parameters, the hyperparameters, and
the map of interest. Regarding experimental data processing, the
present paper follows~\citep{Orieux12} on inversion for the SPIRE
instrument onboard Herschel, which requires the hyperparameters to be
fixed by hand and the instrument parameters to be known. The proposed
method can automatically tune these parameters and may permit the
systematic and automatic processing of large information streams
coming from present and future space-based instruments (e.g. Herschel,
Planck, JWST, etc.).

The paper is structured as follows. Section~\ref{Sec:ModeleDirect}
introduces the notation and sets out the
problem. Section~\ref{Sec:BayesInversion} presents the inversion
method: it introduces the \prior densities and leads to the \post
density. Section~\ref{Sec:EstimAlgo} describes the computing method
based on Markov Chain Monte-Carlo (Gibbs) stochastic sampling
algorithms. The work is essentially evaluated on simulated
observations and on a first set of real observations in the context of
the SPIRE instrument onboard Herschel. The results are presented in
Section~\ref{Sec:Resultats}. Finally, some conclusions and
perspectives are provided in Section~\ref{Sec:Conclusion}.

\section{Notation, instrument, and map models}
\label{Sec:ModeleDirect}

To produce accurate and reliable maps, the inversion must exploit a
description that represents the acquisition process as faithfully as
possible. In this sense, the instrument model
\begin{itemize}
\item is based on a map of the sky noted $\Ciel$, which is naturally a
  function of continuous spatial variables $(\alpha,\beta)\in\eR^2$
  (and possibly a spectral variable $\lambda\in\eR_+$),

\item and accurately describes the formation of a set of $N$ discrete
  observations grouped together in a vector $\yb\in\eR^N$.
\end{itemize}

A general description of the map of the sky as a function of
continuous spatial variables can be written starting from a basic
function~$\psi$ by combination and regular shifting:
\begin{equation}\label{Eq:DecompoCiel}
  \Ciel(\alpha,\beta) = \sum_{ij} x_{ij} ~ \psi\left(\alpha - i~\Ta, \beta - j~\Tb\right).
\end{equation}
The function $\psi$ must be chosen so that this decomposition can
describe the maps of interest and is easy to handle. It may be, among
other choices, a pixel indicator, a cardinal sine function, or a
wavelet (although in the last case the function $\psi$ and the
coefficients also depend on a scaling parameter). Whatever the choice,
the map of interest is finally represented by its coefficients
$x_{ij}$, the number of which is arbitrarily large and collected in
$\xb\in\eR^M$ in what follows. In practice, we choose the Gaussian
family as this greatly simplifies the (theoretical and numerical)
calculations of the model outputs, including for complex
models~\citep{Orieux12,Rodet08}.

The presented work is quite generic in the sense that it is not \aprio
attached to a specific instrument. It deals with a general linear
instrument model that describes, at least to a fair approximation, the
physics of the processes in play: optics, electrics, and
thermodynamics. It also includes the passage from a continuous
physical reality to a finite number of discrete observations. The
instrument is then described by
\begin{equation}
  \label{Eq:DirectTotal}
  \yb = \Ab_\etab \xb + \nb ,
\end{equation}
i.e. a general linear model \wrt $\xb$ (a special case of which is the
convolutive model). This model shows the instrument parameters
$\etab\in\eR^K$ that define the form of the instrument response. The
component $\nb=\yb-\Ab_\etab \xb$ represents the measuring and
modelling errors additively.  For the SPIRE instrument
\citep{Griffin10} of the Herschel space observatory \citep{Pilbratt10}
launched in May 2009, the paper of \citep{Orieux12} gives the details
of the instrument model construction. The results of
Section~\ref{Sec:Resultats} are based on this instrument.

\section{Probabilistic models and inversion}
\label{Sec:BayesInversion}

The proposed inversion is developed in the framework of Bayesian
statistics. It relies on the \post density $p(\xb,\xib,\etab|\yb)$ for
the unknown quantities $\xb$ (image), $\xib$ (hyperparameters), and
$\etab$ (instrument parameters) given $\yb$ (observations). This
density brings together the information about the unknowns in the
sense that it attaches more or less confidence to each value of the
triplet $(\xb,\xib,\etab)$. A summary of this density in the form of a
mean and a standard deviation will provide (1) a point estimate (the
\post mean) for the map of interest and the parameters, and (2) an
indication of the associated uncertainty (the \post standard
deviation).

\begin{remark}\label{RQ:MMSE}
  In statistical terms~\citep{Robert05}, the \post mean is an optimal
  estimator. More precisely, of all the possible estimators (whether
  Bayesian or not, empirical or not, a computation code, etc.), the
  \post mean yields the minimum mean square error (MMSE)\footnote{The
    mean square error is the expected value of the squared norm of the
    difference between estimated value and true value. The expectation
    is under the distribution of the observation and the unknown. The
    MSE is the sum of the variance and the squared bias of the
    estimator (under the distribution of the observation and the
    unknown).}. Regarding first-order statistics, this estimator has,
  moreover, a zero mean bias.
\end{remark}

The \post density is deduced as the ratio of the joint density for all
considered quantities $p(\xb,\xib,\etab,\yb)$ and the marginal density
for the observations $p(\yb)$ by application of Bayes' rule
\begin{equation}
  \label{Eq:PosteriorGenerique}
  p(\xb,\xib,\etab | \yb) = \frac{p(\xb,\xib,\etab,\yb)}{p(\yb)}.
\end{equation}
Seen as a function of the unknowns $(\xb,\xib,\etab)$, this \post
density is proportional to the joint density:
\begin{equation}
  p(\xb,\xib,\etab | \yb) \propto p(\xb,\xib,\etab,\yb) .
\end{equation}
This joint density is essential as all the other densities (marginal,
conditional, \prior, \post, etc.) can be deduced from it. It can be
factorised in various forms and, in preparatio for the developments to
follow, we write
\begin{align} \label{Eq:JointeGenerique}
  p(\xb,\xib,\etab,\yb)	&= p(\yb|\xb,\xib,\etab) p(\xb,\xib,\etab) \nonumber\\
  &= p(\yb|\xb,\xib,\etab) p(\xb|\xib,\etab) p(\xib,\etab) \nonumber\\
  &= p(\yb|\xb,\xib,\etab) p(\xb|\xib) p(\xib) p(\etab) ,
\end{align}
including the fact that (1) the hyperparameters $\xib$ and the
instrument parameters $\etab$ are \aprio independent and (2) the
object $\xb$ and the instrument parameters $\etab$ are also \aprio
independent.

The different probability densities will be defined in the following
sections according to the information available on each set of
variables and according to practical concerns about dealing with the
probability densities and numerical computation time.

\subsection{Modelling of errors and
  likelihood} \label{sec:modelisation-de-la}

The factor $p(\yb|\xb,\xib,\etab)$ in Eq.~\eqref{Eq:JointeGenerique}
is the density for the observations $\yb$ given the map $\xb$, the
instrument parameters $\etab$, and the hyperparameters~$\xib$,
i.e. the likelihood of the unknowns attached to the observations.

Given Eq.~\eqref{Eq:DirectTotal}, the construction of this likelihood
is based on the model for the error $\nb$. The analysis developed in
this paper is essentially founded on its mean $m_\nb$ and its
covariance matrix $\Sigmab_\nb$, and the proposed model is Gaussian:
\begin{equation}
  \label{Eq:NoiseLaw}
  \nb \sim \Ncal(m_\nb, \Sigmab_\nb ).
\end{equation}
The choice of the Gaussian model is also justified via information
property: based on the sole information of finite mean and covariance,
the Gaussian density is the model that introduces the least
information~\citep{Kass96}. This property is also mentioned as a
maximum entropy property of the Gaussian density.

$m_\nb$ is a scalar that models a possible non-zero mean for the
noise, such as an offset (in the proposed numerical evaluation of
Section~\ref{Sec:Resultats}, one offset for each bolometer is
introduced).
Regarding the covariance matrix, to lighten the notation, we set
$\Sigmab_\nb^{-1} = \gamma_\nb \Pib_\nb$: $\gamma_\nb$ is a scale
factor (called precision, homogeneous to an inverse variance) and
$\Pib_\nb$ contains the structure itself. For a stationary model
$\Pib_\nb^{-1}$ has a \Toeplitz structure, for an auto-regressive
model $\Pib_\nb$ is a band matrix, for a white model $\Pib_\nb$ is
diagonal, for a white and stationary model $\Pib_\nb=\Ib$, the
identity matrix. In the developments below, the structure of
$\Pib_\nb$ is given while the scale factor $\gamma_\nb$ and the mean
$m_\nb$ are unknown and included in the vector $\xib$. The results in
Section~\ref{Sec:Resultats} are presented for the case $\Pib_\nb =
\Ib$; hence $\gamma_\nb$ is the inverse of the noise power.

\begin{remark}
  The proposed developments account for characteristics of the error
  $\nb$ that may differ from channel to channel, sensor to
  sensor, etc. This will be the case in Section~\ref{Sec:Resultats}: a
  mean and power of the noise will be assigned and estimated for each
  bolometer.
\end{remark}

As the error $\nb$ is Gaussian and additive (Eqs.~\eqref{Eq:NoiseLaw}
and~\eqref{Eq:DirectTotal}), the vector of observations $\yb$, given
$\xb,\xib,\etab$, is also Gaussian
\begin{equation*}
  \yb|\xb,\xib,\etab\sim \Ncal (\mb_{\yb|*}, \Sigmab_{\yb|*})
\end{equation*}
with mean
\begin{equation}
  \label{Eq:MoyCovVraisemblance}
  \mb_{\yb|*} =\Ab_\etab \xb + m_\nb
\end{equation}
and with the same covariance as $\nb$: $\Sigmab_{\yb|*} =
\Sigmab_\nb$.  So, the likelihood of the unknowns attached to the
observations reads
\begin{multline}
  \label{Eq:Vraisemblance}
  p(\yb|\xb,\xib,\etab) = (2\pi)^{-N/2} \gamma_\nb^{N/2}
  \det\big[\Pib_\nb\big]^{1/2}\\ \Exp{-\frac{1}{2} \gamma_\nb (\yb -
    \mb_{\yb|*})^t \Pib_\nb (\yb - \mb_{\yb|*})}.
\end{multline}
It includes the information provided by the observations as the
transform of a map $\xb$ by the instrument, taking its parameters
$\etab$ and the noise parameters $\gamma_\nb$ and $m_\nb$ into
consideration.

\subsection{Prior density for the map and spatial
  regularity}\label{Sec:PriorCiel}

The aim of this section is to introduce a \prior density $p(\xb
|\xib)$ for the unknown map coefficients $\xb$ based upon available
information about the map $\Ciel$. The present work is mainly devoted
to extended emissions. From a spatial standpoint, such maps are
relatively regular, i.e. they involve positive correlation. From the
spectral standpoint, the power is mainly located at relatively low
frequencies. The Gaussian density includes these second-order
properties in a simple way. This choice can also be justified based on
a maximum entropy principle. Its main interest here is to result in a
linear processing method. It is written in the form
\begin{multline}
  \label{Eq:PriorObjet}
  p(\xb | \gamma_\xb) = (2 \pi)^{-M/2} \gamma_\xb^{M/2}
  \det\big[\Pib_\xb\big]^{1/2} \\
  \Exp{-\frac{1}{2} \gamma_\xb \xb^t \Pib_\xb \xb},
\end{multline}
where $\gamma_\xb$ is a precision parameter (homogeneous to an
inverse-variance that controls the regularity strength) and $\Pib_\xb$
is a precision matrix (homogeneous to an inverse-covariance matrix
that controls the regularity structure). When the precision
$\gamma_\xb$ is low (strong \prior variance), the regularity
information is weakly taken into account. Conversely, when the
precision $\gamma_\xb$ is high (weak \prior variance), the
penalization of non-regular maps is high, i.e. the regularity is
strongly imposed.

The subsequent developments are devoted to the design of $\Pib_\xb$ to
account for the desired regularity of the map.  A simple regularity
measure $\Rc\cro{\Ciel}$ of the map $\Xc$ is the energy of some of its
derivatives. These derivatives can address the spatial variables
$(\alpha,\beta)$ separately, can rely on cross derivatives and can
intervene at various orders. This is the classical
Philipps-Twomey-Thikonov penalization idea~\citep{Tikhonov77}. It can
also embed directional derivatives or any differential
operator~\citep{Mallat08}. In the simplest and natural case, we
choose
\begin{equation*}
  \Rc\cro{\Ciel} = \left\| \frac{\partial \Ciel}{\partial \alpha} \right\|^2 +
  \left\| \frac{\partial \Ciel}{\partial \beta} \right\|^2 ,
\end{equation*}
where $\left\| u \right\|$ is the standard function
norm\,\footnote{The function squared norm is defined by $\left\| u
    \right\|^2=\iint u(\alpha,\beta)^2\, \dD \alpha \,\dD \beta$.}.
Given the decomposition~\eqref{Eq:DecompoCiel}, it is easy to
establish the partial derivatives of $\Ciel$ from the derivatives of
$\psi$. In the direction $\alpha$, by noting $\psi_{\alpha}'
= \partial\psi / \partial \alpha$, we have
\begin{multline*}
  \left\| \frac{\partial \Ciel}{\partial \alpha} \right\|^2 =
  \sum_{ij~i'j'} x_{ij} ~ x_{i'j'} \\ \int_{\eR^2}
  \psi_{\alpha}'\Big(\alpha - i'~\Ta, \beta - j'~\Tb\Big) \\
  \psi_{\alpha}'\Big(\alpha - i~\Ta, \beta - j~\Tb\Big)\, \dD \alpha\,
  \dD \beta,
\end{multline*}
which brings out the autocorrelation $\Psi_{\alpha} = \psi_{\alpha}'
\conv \psi_{\alpha}'$ of the derivative of $\psi$. We then have a
quadratic form in~$\xb$
\begin{multline*}
  \left\| \frac{\partial \Ciel}{\partial \alpha} \right\|^2 =
  \sum_{ij~i'j'} x_{ij} ~ x_{i'j'} \\ \Psi_{\alpha} \left[ (i'- i)~\Ta ,
    (j'- j)~\Tb \right] = \xb^t \Psib_\alpha \xb.
\end{multline*}
As the coefficients $\Psi_{\alpha} \left[ (i'- i)~\Ta , (j'- j)~\Tb
\right]$ depend only on the difference between indices, the matrix
$\Psib_\alpha$ has a \Toeplitz structure and the computations amounts
to a discrete convolution that can be efficiently implemented by the
use of fast Fourier transform (FFT).  Finally, by performing the same
development in the $\beta$ dimension, a global quadratic norm appears:
$\Rc\cro{\Ciel} = \xb^t (\Psib_\alpha+ \Psib_\beta) \xb$ and designs
the precision matrix $\Pib_\xb=\Psib_\alpha+ \Psib_\beta$. For more
details and for a spectral interpretation, see Section~2.1 and
Appendix~A of~\citep{Orieux12}.

\subsection{Prior distribution for hyperparameters
  (hyperprior)} \label{Sec:PriorHyper}

The hyperparameters are the unknown parameters of the densities for
the error and for map
Eqs.~\eqref{Eq:Vraisemblance}-\eqref{Eq:PriorObjet} and they are
collected in the vector $\xib=[m_\nb, \gamma_\nb, \gamma_\xb]$.  It
has been said that $\gamma_\xb$ and $\gamma_\nb$ are the precisions
(scale parameters) and $m_\nb$ is a mean (position parameter) of
Gaussian densities.

The choice of the \prior distributions for these hyperparameters is
driven by two requirements: (i) little information is available \aprio
on their values and their relations and (ii) the chosen distributions
must lead to efficient algorithms (see section~\ref{Sec:HyperSample}).
Following this line of thought, we choose a prior distribution
determined by Jeffreys' principle\footnote{It yields a non-informative
  prior distribution based on a key feature that it is invariant under
  reparameterization. It is deduced as the determinant of the Fisher
  information matrix.}: $p(\gamma) = 1/\gamma$ for $\gamma_\xb$ and
$\gamma_\nb$ and $p(m_{\nb})=1$.  Moreover, regarding the triplet of
hyperparameters $\xib=[m_\nb, \gamma_\xb, \gamma_\nb]$, they are
modelled as independent variables, since no information is available
about their eventual relations.  Finally
\begin{equation}
  \label{Eq:PriorJeffreys}
  p(m_\nb, \gamma_\xb, \gamma_\nb) = 1 / \gamma_\xb \gamma_\nb,
\end{equation}
has two advantages
\begin{enumerate}
\item First of all, the \post conditional densities for $\gamma_\xb$
  and $\gamma_\nb$ (resp. for $m_\nb$), as shown in
  section~\ref{Sec:HyperSample}, will be gamma densities
  (resp. Gaussian density), which will make the implementation easier.

\item This prior distribution is non-informative (which introduces a
  minimum of information on the value of the hyperparameters) in the
  sense that it is invariant by certain parameterization
  changes~\citep{Robert05,Kass96}.
\end{enumerate}

\subsection{Prior density for the instrument
  parameters} \label{Sec:PriorInstrument}

The instrument parameter $\etab$ operates in a complex nonlinear way
in the description of the observations. In consequence, whatever the
\prior density, the conditional \post density for $\etab$ (see
section~\ref{Sec:InstruSample}) will not have a standard form.  The
choice is thus purely oriented by the information on the instruments
and the question that arises concerns the encoding of the available
information in the form of a probability density. If we have no
information except a minimum and a maximum value for a given
parameter, the choice of a uniform density over the interval is a
reasonable one. If we have a nominal value with an associated
uncertainty and no other information, the most suitable choice is a
Gaussian density. The rest of the development is valid whatever the
choice, and we consider the Gaussian case in the following.

In addition, having no information available about possible links
among the various parameters, we take it that the parameters are,
\aprio, independent and thus
\begin{equation}
  \label{Eq:PriorInstru}
  p(\etab) = \prod_{k=1}^K (2\pi \rho_k)^{-1/2}
  \Exp{ - \frac{ (\eta_k - \mu_k)^2 }{ 2\rho_k } }.
\end{equation}
In practice and for the first results presented in
Section~\ref{Sec:Resultats}, the means $\mu_k$ and variances $\rho_k$
were taken from the SPIRE observer manual or were fixed ad-hoc at
plausible values.

\begin{figure}[tb]
  \begin{center}
    \begin{tikzpicture}[scale=0.8]
      \node[draw, circle]{$~\yb~$} [<-,grow=up] child
      {node[draw,circle] {$\etab$} child {node[draw] {$~\rho~$}} child
        {node[draw] {$\mu$}} } child {node[draw,circle] {$m_\nb$}}
      child {node[draw] {$\Pib_\nb$}} child {node[draw,circle]
        {$\gamma_\nb$}} child {node[draw,circle] {$~\xb~$} child
        {node[draw,circle] {$\gamma_\xb$}} child {node[draw]
          {$\Pib_\xb$}} };
    \end{tikzpicture}
  \end{center}
  \caption{Graphical dependency representation (hierarchical
    structure). The round (square) nodes correspond to unknown (fixed)
    quantities. The directions of the arrows indicate the
    dependencies.\label{fig:ModelGraphique}}
\end{figure}
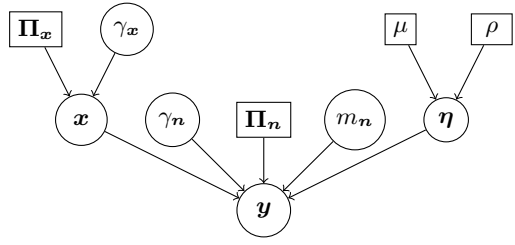

\subsection{\Post density: histograms, mean and standard
  deviation} \label{Sec:Posterior}

The \post density~\eqref{Eq:PosteriorGenerique} for all the unknowns
$\xb,\xib,\etab$, is deduced from the joint
density~\eqref{Eq:JointeGenerique} for all quantities concerned as
follows:
\begin{equation*}
  p(\xb,\xib,\etab | \yb) \propto p(\yb|\xb,\xib,\etab) p(\xb|\xib)
  p(\xib) p(\etab) .
\end{equation*}
In this expression,
\begin{itemize}
\item the density $p(\xb|\gamma_\xb)$ for the unknown map
  $\xb\in\eR^M$ is Gaussian (Eq.~\eqref{Eq:PriorObjet});
\item the distribution $p(\xib)$ (for $\xib=[m_\nb, \gamma_\nb,
  \gamma_\xb]$) is a Jeffreys' distribution
  (Eq.~\eqref{Eq:PriorJeffreys});
\item the instrument parameters $\etab\in\eR^K$ is modelled by a
  Gaussian density $p(\etab)$ (Eq.~\eqref{Eq:PriorInstru});
\item the density $p(\yb|\xb,\xib,\etab)$ for the observations $\yb
  \in \eR^N$ given the rest of the variables (i.e. likelihood) is a
  Gaussian density (Eq.~\eqref{Eq:Vraisemblance}) and it is a function
  of $\xb$ through $\mb_{\yb|*}$ given
  by~Eq.~\eqref{Eq:MoyCovVraisemblance}.
\end{itemize}

Finally, from equations~\eqref{Eq:Vraisemblance},
\eqref{Eq:PriorObjet}, \eqref{Eq:PriorJeffreys},
and~\eqref{Eq:PriorInstru}, the \post density can be written
\begin{multline}
  \label{Eq:PosteriorTotale}
  p(\xb,\xib,\etab | \yb) \propto \gamma_\xb^{M/2 - 1} ~
  \gamma_\nb^{N/2 - 1} \prod_{k=1}^K ( 2\pi \rho_k )^{-1} \\
  \Exp{ -\frac{1}{2} \frac{ (\eta_k - \mu_k)^2 }{ \rho_k }}
  \Exp{-\frac{1}{2} ~\gamma_\xb~ \xb^t \Pib_\xb \xb} \\
  \Exp{-\frac{1}{2} \gamma_\nb (\yb - \mb_{\yb|*})^t \Pib_\nb (\yb -
    \mb_{\yb|*})}.
\end{multline}

This density brings together all information about the unknowns, and
the estimators and algorithms presented below are entirely based on
it. However, it is too complex to be analyzed directly as a whole and
the difficulty stems from (i) the dimension of $\xb$ (of size $M\sim
10^5$ in practice) and (ii) the joint presence of other parameters
(hyperparameters $\xib$ and instrument parameters $\etab$). Moreover,
for the latter in particular, the dependence is complicated and cannot
be identified with a standard form. The proposed approach is to
explore the \post density by means of stochastic
sampling~\citep{Robert04,Gilks96}. The idea is to produce a set of
samples $\xb^{(q)}, \xib^{(q)} , \etab^{(q)}$, for $q=1,2,\dots,Q$,
drawn at random under the \post density. It is then possible, for
example, to deduce histograms that approximate marginal densities,
together with means and standard deviations. This strategy is by no
means new but interest in its practical use has revived in recent
years as new forms of algorithms have been developed and computer
power has increased.

Concerning the estimates themselves for the map, the hyperparameters
and the instrument parameters, we choose the \post mean (PM), as
indicated at the beginning of Section~\ref{Sec:BayesInversion} (see
also Remark~\ref{RQ:MMSE}). We will also look at the dispersion around
the mean through the \post standard deviation (PSD) and the links
among components through \post correlations.  Using the set of samples
$\xb^{(q)}, \xib^{(q)} , \etab^{(q)}$, for $q=1,2,\dots,Q$, the
posterior mean $\mub_{\PD}$ and the posterior covariance matrix
$\Gammab_{\PD}$ are computed by
\begin{gather}
  \mub_{\PD} \approx \frac{1}{Q} \sum_{q=1}^{Q} \bar\xb^{(q)} \label{eq:MoyEmp} \\
  \Gammab_{\PD} \approx \frac{1}{Q} \sum_{q=1}^{Q} \left(
    \bar\xb^{(q)}- \mub_{\PD} \right) \left(\bar\xb^{(q)} -
    \mub_{\PD}\right)^t, \label{eq:CorEmp}
\end{gather}
where $\bar\xb$ denotes the column concatenation
$\bar\xb=[\xb;\xib;\etab]$. Practically, it is not possible to compute
the entire covariance $\Gammab_{\PD}$, but it is possible to compute
its diagonal elements to characterize the marginal errors for each
component (each pixel, hyperparameters, instrument parameters) and to
compute a few nondiagonal elements to measure the correlations between
components.

\section{Exploration of \post density and the computation
  algorithm} \label{Sec:EstimAlgo} \label{Sec:TechnExplor}

We have introduced an instrument model and various probability
densities to define the \post density that brings together the
information on the map and the parameters (hyperparameters and
instrument parameters). We have also defined the \post mean (PM) as an
estimate and the \post standard deviation (PSD) as a measure of the
uncertainty.  We have then introduced the idea of computations via
stochastic sampling. The developments in the present section concern
the algorithm for computing these samples.

\begin{table}[tb]
  \caption{Gibbs algorithm\label{Algo:Gibbs}}
  \begin{algorithmic}[0]
    \State Initialize $\xb^{(0)}, \xib^{(0)}, \etab^{(0)}$

    \For{$q=1,2,\dots$}

    \State (1) sample $\xb^{(q)}$ under $p( \xb | \xib^{(q-1)},
    \etab^{(q-1)}, \yb)$

    \State (2) sample $\xib^{(q)}$ under $p( \xib | \xb^{(q)},
    \etab^{(q-1)}, \yb)$

    \State (3) sample $\etab^{(q)}$ under $p( \etab | \xb^{(q)},
    \xib^{(q)}, \yb)$
    \EndFor
  \end{algorithmic}
\end{table}

The production of samples of the \post density for the set
$(\xb,\xib,\etab)$ is not possible directly because of the complexity
of the density. We therefore use a Gibbs
algorithm~\citep{Robert04,Gilks96}, which breaks the problem down into
three simpler subproblems: sampling $\xb$, $\xib$, and $\etab$
separately.  This is an iterative algorithm, described in
Tab.~\ref{Algo:Gibbs}: each variable $\xb$, $\xib$, and $\etab$ is
drawn under its conditional \post density given the current value of
the other two variables. For each of the three steps, this conditional
\post density can be deduced directly (up to a multiplying factor)
from the \post density \eqref{Eq:PosteriorTotale}: all we have to do
is to keep only the factors depending on the variable of interest.
This algorithm is a Markov chain Monte-Carlo (MCMC)
algorithm~\citep{Robert04,Gilks96} and is known to give (after a
certain time, called the burn-in time) samples under the \post
density.

The conditional density of the map coefficients $\xb$ (Step~(1),
Tab.~\ref{Algo:Gibbs}) is Gaussian
(Section~\ref{Sec:ObjectSample}). For the precisions $\xib$ (Step~(2),
Tab.~\ref{Algo:Gibbs}), the conditional densities are gamma densities
(Section~\ref{Sec:HyperSample}).  They will be sampled using standard
existing numerical routines (e.g. in Matlab). In contrast, the
conditional density of the instrument parameters $\etab$ (Step~(3),
Tab.~\ref{Algo:Gibbs}) has a much more complex nonstandard form, so
that it cannot be directly sampled by existing routines. To overcome
this difficulty, sampling was carried out by means of a
Metropolis-Hastings step (Section~\ref{Sec:InstruSample}).

\subsection{Map sampling} \label{Sec:ObjectSample}

The density for the map $\xb$ conditionally on the other variables is
deduced from~\eqref{Eq:PosteriorTotale} by extracting the factors
depending on $\xb$:
\begin{multline} \label{eq:PostCondObjet}
  p(\xb|\yb,\gamma_\xb,\gamma_\nb,\etab) \propto \exp \Big[
  \frac{1}{2} \gamma_\xb \xb^t \Pib_\xb \xb \,+ \\ \gamma_\nb (\yb -
  \mb_{\yb|*})^t \Pib_\nb (\yb - \mb_{\yb|*}) \Big].
\end{multline}
Considering the expression for $\mb_{\yb|*}$ given by
Eq.~\eqref{Eq:MoyCovVraisemblance}, the argument of this exponential
is quadratic in $\xb$. We deduce that we have a Gaussian density and,
by rearranging the argument, we can determine the covariance and the
mean
\begin{equation}\label{Eq:CondPostObjetCOV}
  \Sigmab_{\xb|*} = \left( \gamma_\nb \Ab_{\etab}^t \Pib_\nb
    \Ab_{\etab} + \gamma_\xb \Pib_\xb \right)^{-1}
\end{equation}
\begin{equation}\label{Eq:CondPostObjetMOY}
  \mb_{\xb|*} = \gamma_\nb\Sigmab_{\xb|*}\Ab_{\etab}^t\Pib_\nb\yb,
\end{equation}

\begin{remark}\label{RQ:SolMCRPosteriorCond}
  For a fixed value of the hyperparameters and the instrument
  parameters, the map $\mb_{\xb|*}$ defined by
  Eqs.~\eqref{Eq:CondPostObjetCOV}-\eqref{Eq:CondPostObjetMOY} is the
  maximizer (and the mean) of the conditional \post
  density~\eqref{eq:PostCondObjet}. This is the regularized
  least-squares solution denoted $\wh\xb(\mu)$ parameterized by
  $\mu=\gx/\gB$. This corresponds to the solution defined in our
  previous paper~\citep{Orieux12}. For a convolutive instrument model,
  it is the Wiener solution (also called Wiener-Hunt
  solution)~\citep{Orieux10a}.
\end{remark}

Step~(1) of Tab.~\ref{Algo:Gibbs} consists of sampling this Gaussian
but this operation is made very difficult by three elements: (i) the
large size of the map, (ii) the correlation introduced by the
instrument model and the \prior density, and (iii) the absence of
structure of the instrument model (invariance, sparse nature). This
problem can be solved using several approaches. For a convolutive
instrument model~\eqref{Eq:DirectTotal}, it is possible to
approximately diagonalise the correlation matrix by FFT, thus
producing a sample for the cost of an
FFT~\citep{Chellappa85,Chellappa92,Geman95,Giovannelli08,Orieux10a}. If
where the inverse of the correlation matrix is sparse, a partially
parallel Gibbs sampler may be particularly
efficient~\citep[Chap.8]{Winkler03}. In the present case, neither the
correlation nor its inverse possess the required properties. A general
solution relies on factorizing the correlation matrix (Cholesky
decomposition, diagonalization, etc.) but the large size of the matrix
($M\times M$ with $M\sim 10^5$) does not permit the required
calculations to be performed here.

The proposed solution consists of constructing a criterion such that
its minimizer is a sample under the desired \post conditional
density. To do this, we perturb the means of the noise component and
of the map component by an additive component with covariance
$\gamma_\xb \Pib_\xb$ and $\gamma_\nb \Pib_\nb$.  A perturbed
regularized least squares criterion is then introduced
\begin{multline*}
  J(\xb) = \gamma_\nb \left( \ybt-\Ab_{\etab}\xb \right)^t \Pib_\nb
  \left( \ybt-\Ab_{\etab}\xb \right) + \\ \gamma_\xb \left( \xb-\mxbt
  \right)^t \Pib_\xb (\xb-\mxbt),
\end{multline*}
and it can be shown (see~\citep{Orieux12b}) that its minimizer
\begin{multline}
  \label{Eq:OptimSample}
  \wt{\xb} = \left( \gamma_\nb  \Ab_{\etab}^t  \Pib_\nb
    \Ab_{\etab} + \gamma_\xb  \Pib_\xb \right)^{-1} \\ ( \gamma_\nb
  \Ab_{\etab}^t\Pib_\nb \ybt + \gamma_\xb  \Pib_\xb \mxbt )
\end{multline}
is Gaussian and does indeed have the correlation and mean defined
by~\eqref{Eq:CondPostObjetCOV} and~\eqref{Eq:CondPostObjetMOY}. This
very powerful result has already been used
by~\citep{Feron06,Orieux12}. In different forms, similar ideas have
been introduced and used by~\citep{Rue01,Rue05,Lalanne01,Tan10}.

\begin{remark}\label{RQ:SolMCRPosteriorCondPerturbe}
  For the non perturbed criterion ($\ybt=\yb$ and $\mxbt=0$), we have
  the regularized least-squares solution
  Eqs.~\eqref{Eq:CondPostObjetCOV}-\eqref{Eq:CondPostObjetMOY}, that
  was mentioned in Remark~\ref{RQ:SolMCRPosteriorCond}.
\end{remark}

\begin{remark}\label{RQ:priormean_and_sampling}
  The approach described in \citep{Orieux12b} involves the sampling of
  the prior density~\eqref{Eq:PriorObjet} that is not properly defined
  here: the matrix $\Pib_\xb$ does not penalise the mean of the map
  (it is of deficient rank). But, for the same reason, the
  solution~\eqref{Eq:OptimSample} does not depend on the mean of the
  realization of the prior density. Therefore, the simulated sample
  can have an arbitrary mean value.
\end{remark}

\subsection{Hyperparameter sampling} \label{Sec:HyperSample}

To determine the \post conditional density for $\gamma_\xb$, we
examine the \post density~\eqref{Eq:PosteriorTotale}, and only keep
the factors where~$\gamma_\xb$ appears, which gives
\begin{equation*}
  p(\gamma_\xb|\yb,\xb,\gamma_\nb,\etab)
  \propto \gamma_\xb^{M/2-1} \Exp{- \frac{\gamma_\xb}{2} || \xb
    ||_{\Pib_\xb}^2 },
\end{equation*}
and we recognize a Gamma density (see Appendix~\ref{Sec:GammaLaw})
\begin{equation}
\label{Eq:GammaXPostCond}
\gamma_\xb \sim \Gcal \left( ~M/2~, ~2/|| \xb ||_{\Pib_\xb}^2 ~\right).\
\end{equation}
Concerning $\gamma_\nb$, we also refer to the \post
density~\eqref{Eq:PosteriorTotale} and find
\begin{equation*}
  p(\gamma_\nb|\yb,\xb,\gamma_\xb,\etab) \propto \gamma_\nb^{N/2 - 1}
  \Exp{-\frac{1}{2} \gamma_\nb ||\yb - \mb_{\yb|*} ||_{\Pib_\nb}^2
  },
\end{equation*}
which is also a Gamma density
\begin{equation}
  \label{Eq:GammaNPostCond}
  \gamma_\nb \sim \Gcal \left(~N/2~ , ~2/||\yb - \mb_{\yb|*}
    ||_{\Pib_\nb}^2 ~\right) .
\end{equation}
For both $\gamma_\xb$ and $\gamma_\nb$ the second parameter of the
Gamma density introduces a quadratic norm (regularity of the map in
Eq.~\eqref{Eq:GammaXPostCond}, and goodness-of-fit in
Eq.~\eqref{Eq:GammaNPostCond}), which can be easily computed.

\begin{remark}\label{RQ:HyperSampleInterprete}
  An intuitive interpretation can be given to these results starting
  from the fact that the mean of the Gamma density is equal to the
  product of its parameters (see Appendix~\ref{Sec:GammaLaw}), here
  $N/\left\|\yb - \mb_{\yb|*} \right \|_{\Pib_\nb}^2$
  for~\eqref{Eq:GammaNPostCond}.  In this sense, the conditional
  posterior mean is the inverse of the empirical variance of the
  residuals. Consequently, when the goodness-of-fit term is small, the
  mean of the density is large and so the sampled value of
  $\gamma_\nb$ is also high reporting a high precision, i.e. a weak
  variance (and vice versa).  THe same holds for the map regularity in
  relation with the mean of the density~\eqref{Eq:GammaXPostCond}
  given by $M/\left\|\xb\right\|_{\Pib_\xb}^2$.  These observations
  support the coherence of the model and reinforce the prior choice
  for these hyperparameters as a convenient one.
\end{remark}

Regarding the mean of the noise, $m_\nb$, it is a scalar whose \post
conditional density is also deduced from the \post
density~\eqref{Eq:PosteriorTotale} and
from~\eqref{Eq:MoyCovVraisemblance}
\begin{equation*}
  p(m_\nb|\yb,\xb,\gamma_\nb,\gamma_\xb,\etab) \propto
  \Exp{-\frac{1}{2} \gamma_\nb (m_\nb - m_r)^2} ,
\end{equation*}
where $m_r$ is the empirical mean of the residuals $\yb - \Ab_\etab
\xb$. We then have a Gaussian density
\begin{equation}\label{Eq:GammaNPostCond}
  m_\nb \sim \Ncal \left(m_r, \gamma_\nb \right).
\end{equation}
In the numerical evaluation of Section~\ref{Sec:Resultats}, one such
mean is estimated for each bolometer.

\begin{remark}
  If we examine the relationships above, we see that
  $p(\gamma_\xb|\yb,\xb,\gamma_\nb,\etab) = p(\gamma_\xb|\xb)$, in
  other words, $\gamma_\xb$ and $(\yb,\gamma_\nb,\etab)$ are
  independent conditionally on $\xb$. Similarly, we note that
  $p(\gamma_\nb|\yb,\xb,\gamma_\xb,\etab) =
  p(\gamma_\nb|\yb,\xb,\etab)$, which means that $\gamma_\nb$ and
  $\gamma_\xb$ are independent conditionally on $(\yb,\xb,\etab)$. In
  addition, $m_\nb$ is independant of $\gamma_\xb$, given
  $\yb,\xb,\gamma_\nb,\etab$.
\end{remark}

\begin{table}[tb]
  \caption{Step of the Metropolis-Hastings sampler, which replaces
    Step~(3) of Tab.~\ref{Algo:Gibbs}. The current sample at step $q$ is
    $\etab^{(q)}$ and it is either replaced or not by the proposed sample
    $\etab^\pD$.\label{Algo:MH}}
  \begin{algorithmic}[0]

    \State (a) Draw a sample $\etab^\pD$ under a proposal
    density 

    \State (b) Compute the acceptation ratio $\rho$ by
    Eq.~\eqref{Eq:AcceptProba}

    \State (c) Replace $\etab^{(q)}$ by $\etab^\pD$
    (i.e. $\etab^{(q+1)}=\etab^\pD$) with the probability
    $\min(1,\rho)$, otherwise keep $\etab^{(q)}$
    (i.e. $\etab^{(q+1)}=\etab^{(q)}$).

  \end{algorithmic}
\end{table}

\subsection{Instrument parameter sampling}
\label{Sec:InstruSample}

The last step (Step~(3) of Tab.~\ref{Algo:Gibbs}) is more complex. As
for the other variables, the \post conditional density can be deduced
from the \post density~\eqref{Eq:PosteriorTotale} by keeping the
factors that bring in $\etab$. There are two of these: the likelihood
and the \prior density, and we thus have
\begin{multline}
  \label{Eq:CondPostEta}
  p(\etab|\yb,\xb,\xib) \propto \prod_{k=1}^K \Exp{ -\frac{1}{2}
    \frac{ (\eta_k - \mu_k)^2 }{ \rho_k }} \\ \Exp{-\frac{1}{2}
    \gamma_\nb (\yb - \Ab_\etab\xb)^t \Pib_\nb (\yb - \Ab_\etab\xb) }.
\end{multline}

However, this is not a usual density, notably because there is no
simple mathematical form to represent the dependence of the
observation \wrt $\etab$. Thus, Step~(3) of Tab.~\ref{Algo:Gibbs}
cannot be carried out directly with standard sampling routines and we
resort to a Metropolis-Hastings step in a random-walk
version~\citep{Robert04,Gilks96} described in Tab.~\ref{Algo:MH}. It
can be briefly explained as follows. Because it is impossible to draw
a sample directly under the conditional \post
density~\eqref{Eq:CondPostEta}, a sample is drawn under another
density (namely the proposal density), but is not systematically
accepted. Acceptance or rejection is also random with a precisely
defined probability (see Eq.~\eqref{Eq:AcceptProba}) to ensure that,
at convergence, we have samples under the target
density~\citep{Robert04,Gilks96}. The algorithm is divided into three
sub-steps summarized in Tab.~\ref{Algo:MH} and detailed here.
\begin{itemize}
\item[(a)]~Draw a proposal $\etab^\pD$ as a perturbation of the
  current value: $\etab^\pD = \etab^{(q)} + \varepsilonb$, deduce the
  instrument matrix $\Ab_{\etab^\pD}$, and the corresponding model
  output $\my^{\pD}=\Ab_{\etab^\pD} \xb^{(q)} + m_\nb$.
\item[(b)]~Compute the acceptation ratio
  \begin{equation}
    \label{Eq:AcceptProba}
    \rho =
    \frac{p(\etab^{\pD}|\yb,\xb,\xib)}{p(\etab^{(q)}|\yb,\xb,\xib)},
\end{equation}
based on the conditionnal posterior law ratio that compares the
goodness-of-fit for the current parameter and the proposed one.
\item[(c)]~Accept or reject the proposal, at random, with probability
  $\min(1,\rho)$. To do so, draw $u$ uniformly in $[0,1]$ and take
  \begin{equation*}
    \etab^{(q+1)} = \left\{
      \begin{array}{ll}
        \etab^{\pD} & \text{if~} u < \text{min}\{1,\rho\} \\
        \etab^{(q)} & \text{otherwise}.
      \end{array}
    \right.
  \end{equation*}

\end{itemize}
These three substeps are inserted instead of Step~(3) of
Tab.~\ref{Algo:Gibbs}.

The algorithm can be explained as follows. Starting with a current
value $\etab^{(q)}$, the algorithm proposes a new value $\etab^{\pD}$
and compares the goodness-of-fit for the two values. When the proposed
value improves the fit, $\rho>1$ and $\etab^{\pD}$ is accepted. When
the proposed value degrades the fit, $\etab^{\pD}$ can be accepted or
rejected, with a probability that is higher or lower, depending on how
weak the degradation is.

\begin{remark}
  There are other more complex (and potentially more efficient)
  approaches for Metropolis-Hastings sampling. In particular, the
  proposal density can be adapted, e.g. directional random
  walk~\citep{Vacar11}). They are not exploited here but are
  considered in the development perspectives.
\end{remark}

\section{Numerical results}\label{Sec:Resultats}

The previous sections presented the approach for building the \post
density and for its exploration by stochastic sampling using a Gibbs
algorithm including a Metropolis-Hastings step. The mean and standard
deviation (SD) of the \post density are numerically computed as
empirical averages based on simulated samples, from
relations~\eqref{eq:MoyEmp} and~\eqref{eq:CorEmp}. The developments
below show the practicability of the proposed method (models, estimate
and algorithm), and provide a first numerical evaluation.

\subsection{Evaluation methodology}

\begin{figure}[htb]
  \centering
  \includegraphics[width=0.3\textwidth]{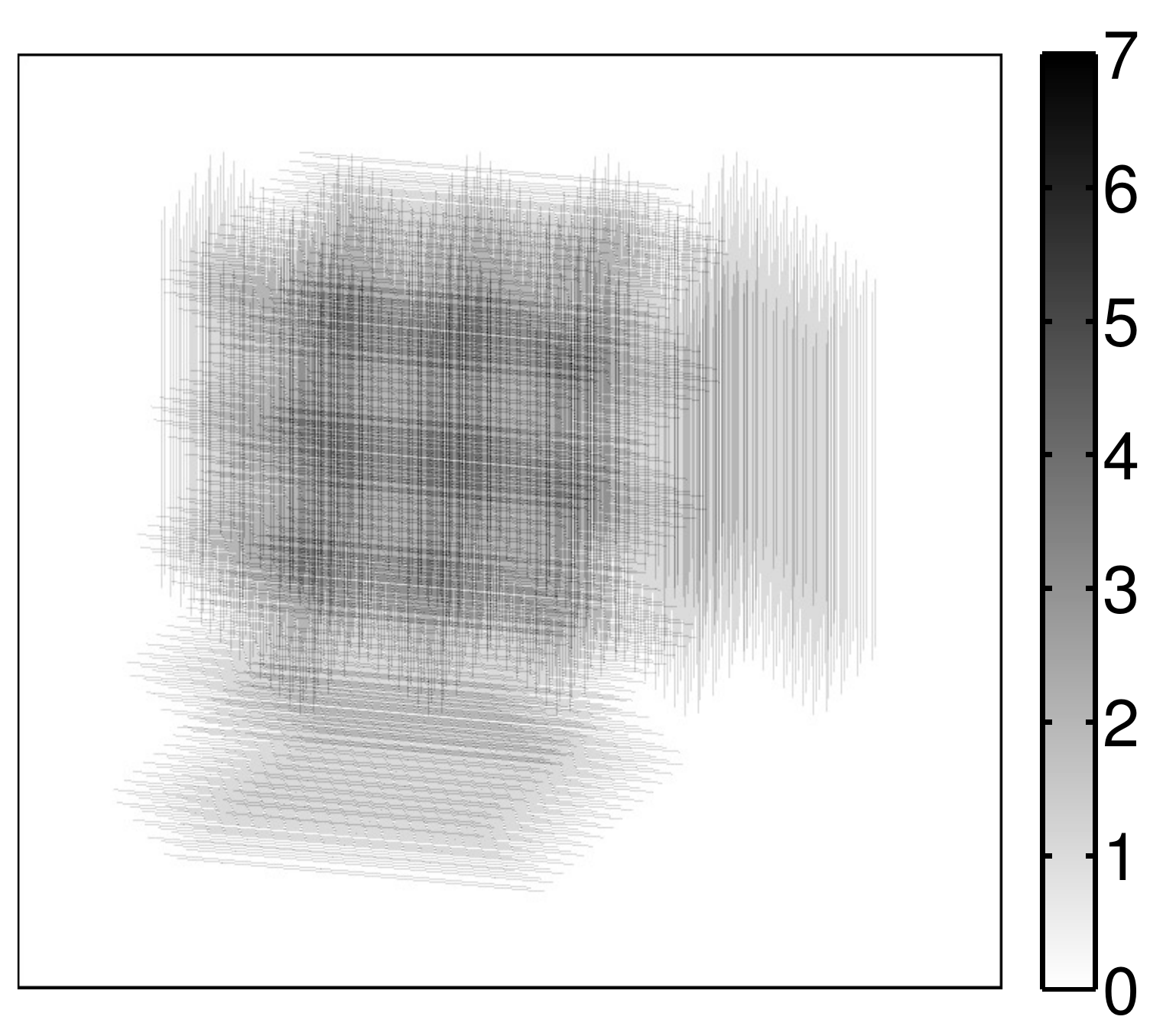}
  \caption{Redundancy/pointing map associated with our experiment of
    five crossed scans.\label{fig:balay}}
\end{figure}

The evaluation is based on the SPIRE instrument \citep{Griffin10} of
the Herschel space observatory \citep{Pilbratt10} launched in May
2009. It focuses on the PMW channel (centred around $350\,\mu$m) and
the \textit{Large Map} protocol in the nominal operating conditions:
scan back and forth with constant speed ($30\,\arcsec/\sec$) over two
almost perpendicular directions~($88^\circ$). The scans are associated
with a high sampling frequency ($\Fe\approx 30$\,Hz) providing
spatially redundant observation and Fig.~\ref{fig:balay} shows the
corresponding redundancy/pointing map.
The spatial shift between basis functions (see
Eq.~\eqref{Eq:DecompoCiel}) is fixed at $\Ta=\Tb=2\,\arcsec$, based on
our earlier work~\citep{Orieux09a,Orieux12}, to obtain the best gain
in resolution without important increase of the computational
cost. The angular size of the reconstructed map is
$20\,\arcmin\times20\,\arcmin$, i.e. a map of $600\times600$
coefficients.
The associated direct model, including the whole acquisition chain
(scanning strategy, mirror, horns, wavelength filters, bolometers,
and electronics) is detailed in our previous paper~\citep{Orieux12} and
represented by Eq.~\eqref{Eq:DirectTotal} of the present paper.

The unsupervised method is assessed based on two synthetic maps of
extended emission (the Galactic Cirrus (Fig.~\ref{fig:true2}) and a
realization of the \prior density Eq.~\eqref{Eq:PriorObjet}) as well
as based on a real observation (reflection nebula NGC\,7023,
Fig.~\ref{fig:hyper-real-data}). The paper also proposes a first
assessment of the unsupervised and myopic approach based on a
synthetic map with broader spectral content (the Galactic Cirrus with
point sources (Fig.~\ref{fig:true2})).

In the simulated cases, a zero-mean white Gaussian noise is added to
the model output. Moreover, in these cases, since the original map
(the ``sky truth'' denoted by $\xb^{*}$) is known, the quality of the
reconstruction (denoted by $\wh\xb$) can be quantified through an
error:
\begin{equation}
  \label{Eq:DefErreurL2}
  \Ec =\sum_{i,j} |\xijstar - \whxij |^2  ~/~  \sum_{i,j} |\xijstar|^2,
\end{equation}
where only coefficients in the observed area are taken into account,
allowing assessments and comparisons between methods.

\subsection{Algorithm behaviour and general comments}

As explained in Section~\ref{Sec:TechnExplor}, the algorithm provides
a series of samples that form a Markov chain for hyperparameters,
instrument parameter and map. The MCMC theory then ensures that it
correctly explores the parameter space and produces a density of
samples reflecting the posterior density. Practically, the algorithm
has been executed for the unsupervised problem as well as for the
unsupervised and myopic problem. It has been run several times (1)
using identical initial conditions and (2) using different initial
conditions. In both cases, the same qualitative and quantitative
behaviour as presented here has been systematically observed.

The computation time takes about one hour for the unsupervised
(nonmyopic) case and about ten hours for the unsupervised and myopic
case. The main computational cost is due to computatingq the
instrument model output given by Eq.~\eqref{Eq:DirectTotal}.

Figs.~\ref{fig:hyperHisto}, \ref{fig:hyper-real-data},
and~\ref{fig:myopicpics} present some typical elements of the
algorithm operation: visualization of progression, convergence phase
(burn-in period), stable phase, etc. The evolution of the chain is
shown for hyperparameters (Figs.~\ref{fig:hyperHisto}
and~\ref{fig:hyper-real-data}) and for instrument parameter
(Fig.~\ref{fig:myopicpics}). It is thus possible to grasp how the
parameter space is explored.

\subsection{Unsupervised approach}

\subsubsection{Assessment of map estimation}\label{sec:etude-sur-le}

\begin{figure*}[htbp]
  \centering

  \subfigure[Proposed map]{\includegraphics[width=0.3\textwidth]
    {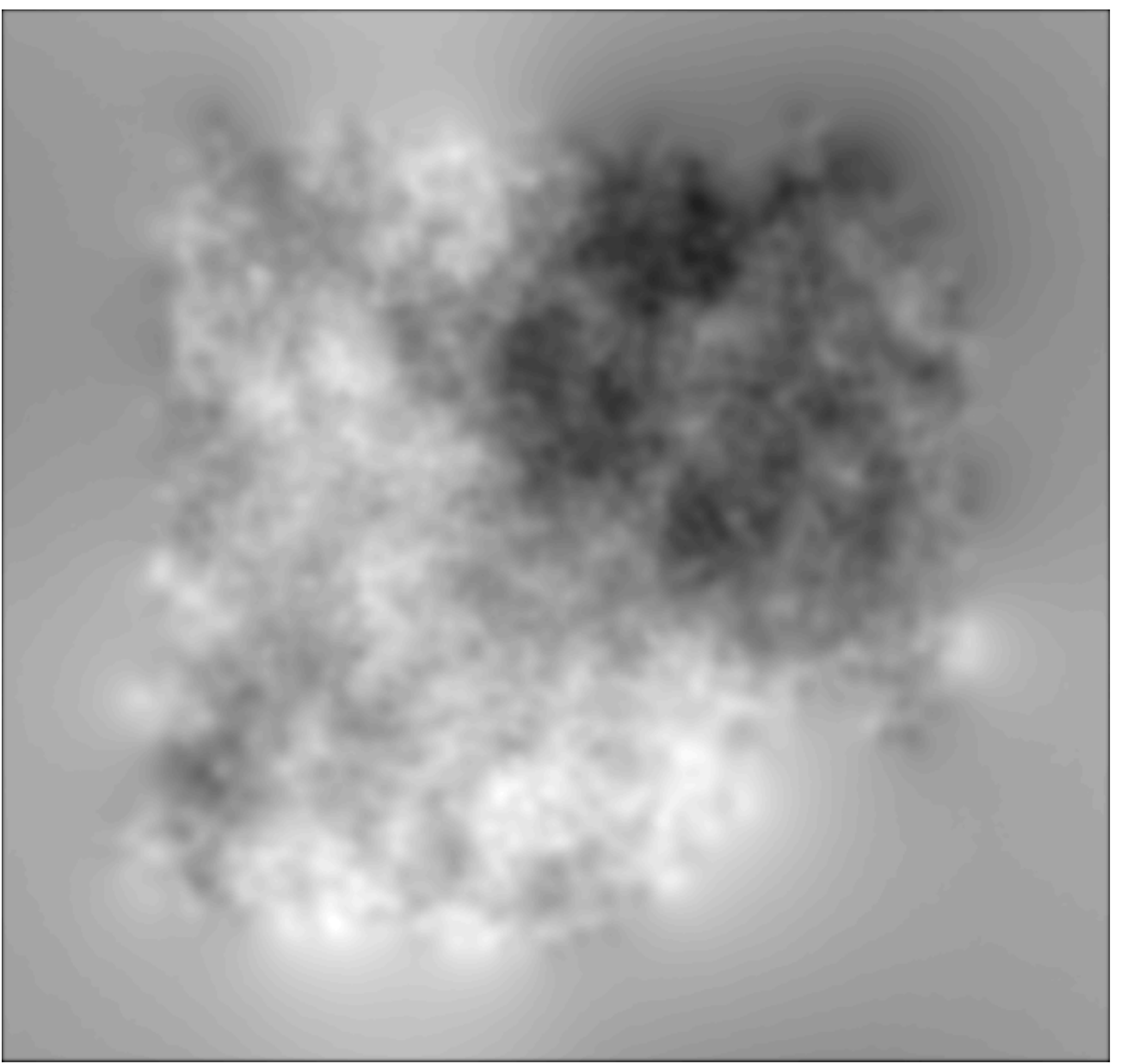}\label{fig:eapCirrusIm}} ~~~~\subfigure[Best
  map]{\includegraphics[width=0.3\textwidth]
    {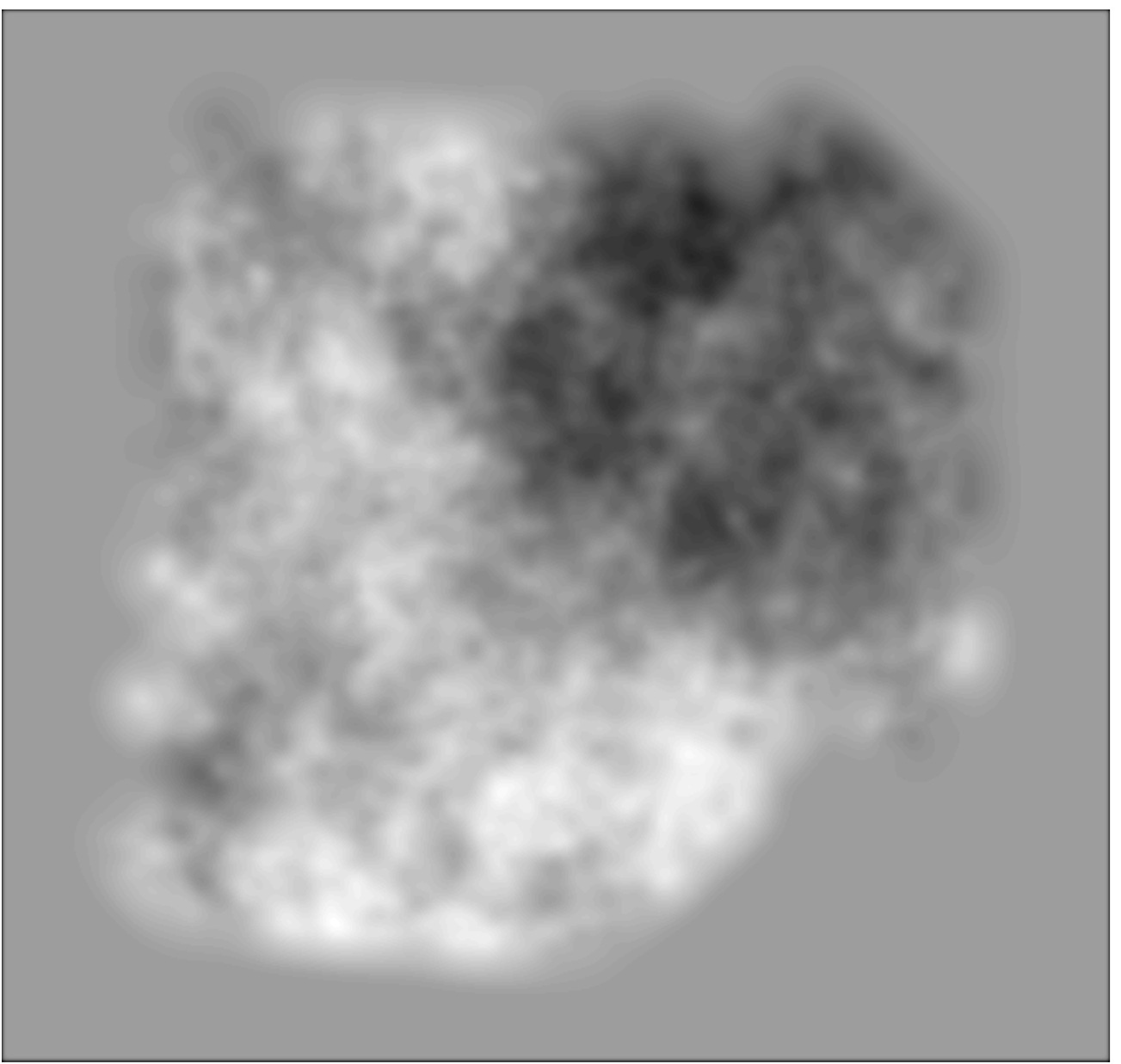}\label{fig:cirrusDiffRegNormal}}
  ~~~~\subfigure[Profile]
  {\includegraphics[width=0.3\textwidth]{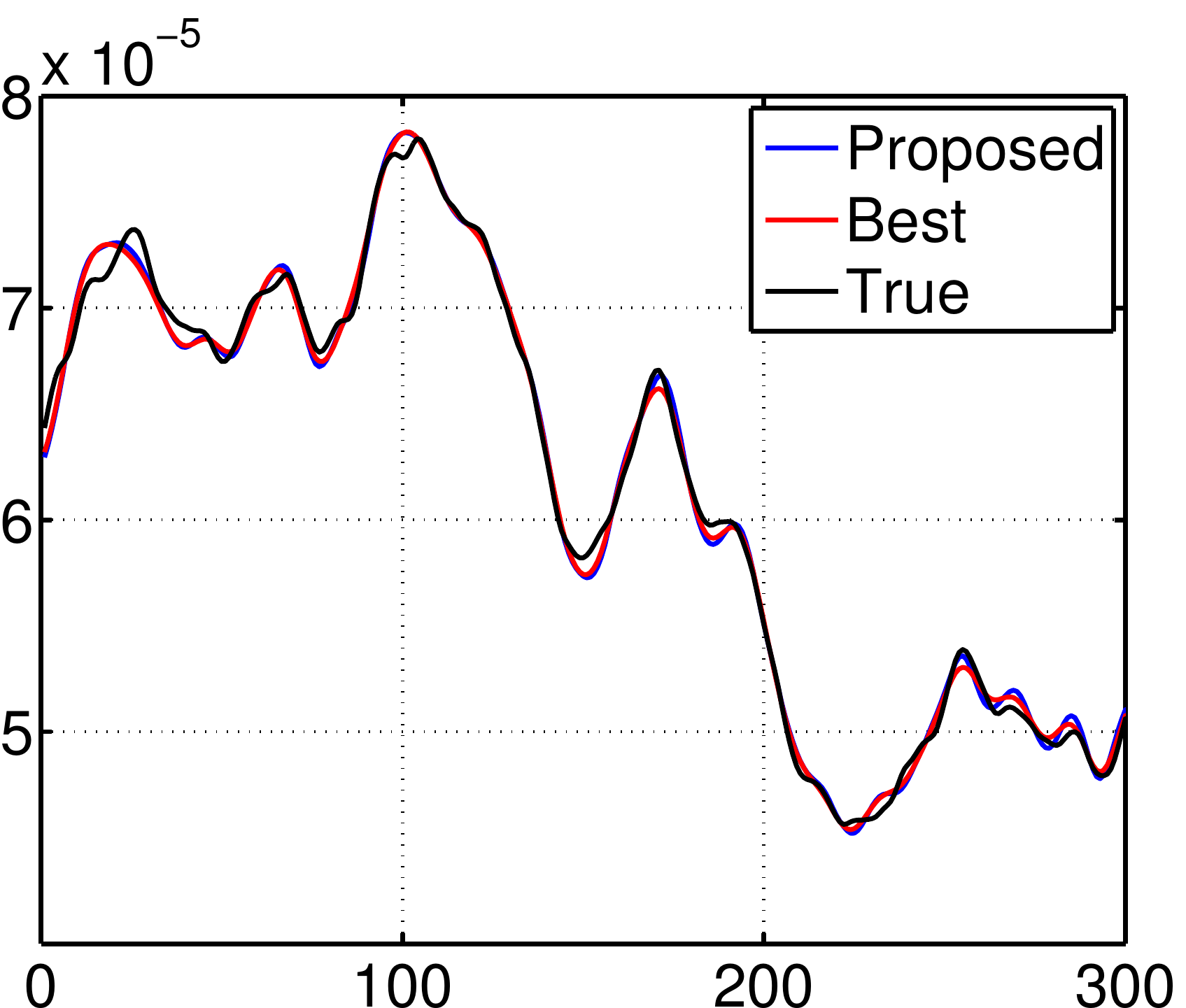}
    \label{fig:sliceEapCirrus}}

  \subfigure[Naive map]{\includegraphics[width=0.3\textwidth]
    {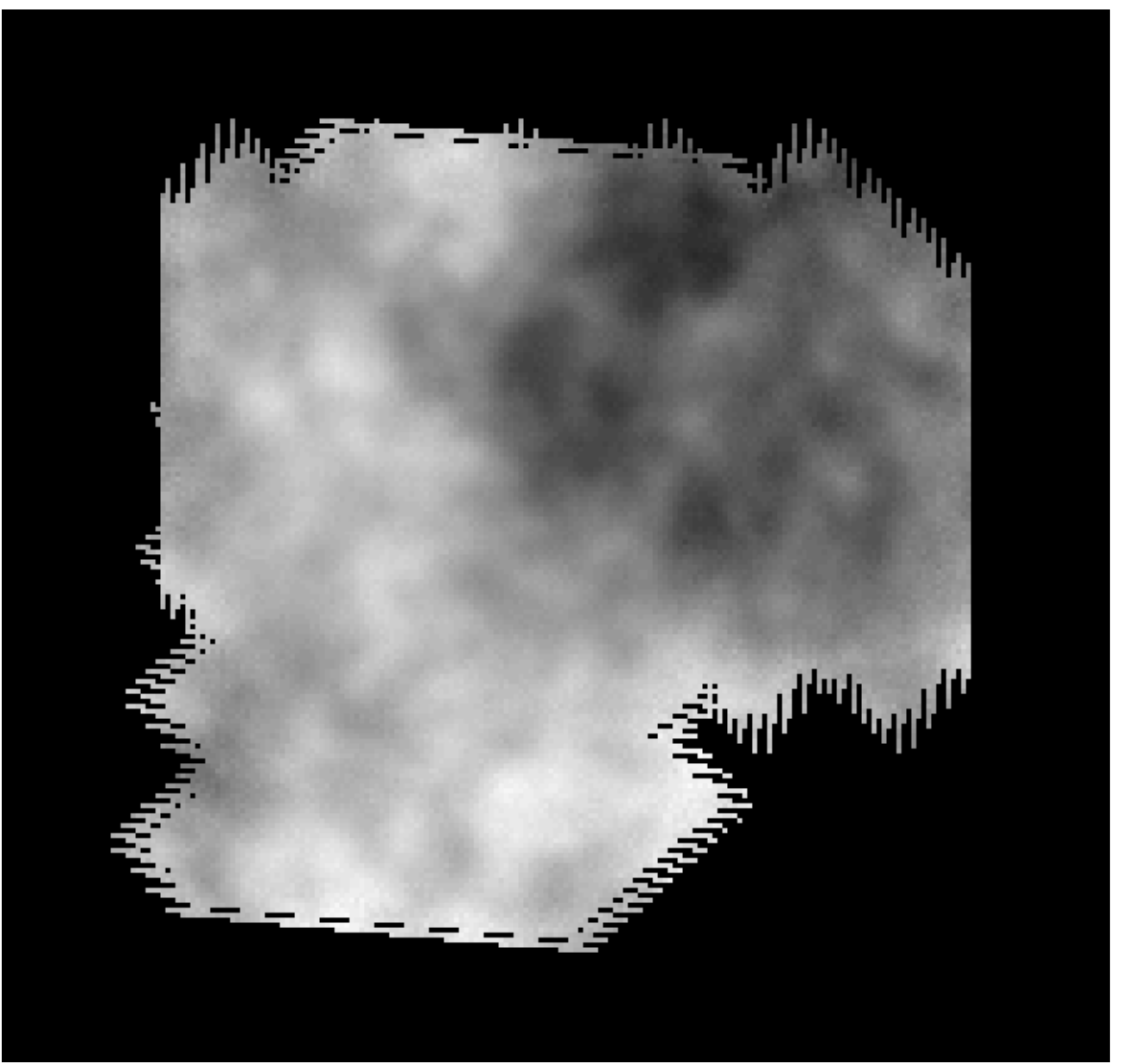}\label{fig:naiveMap}} ~~~~\subfigure[True
  map]{\includegraphics[width=0.3\textwidth]
    {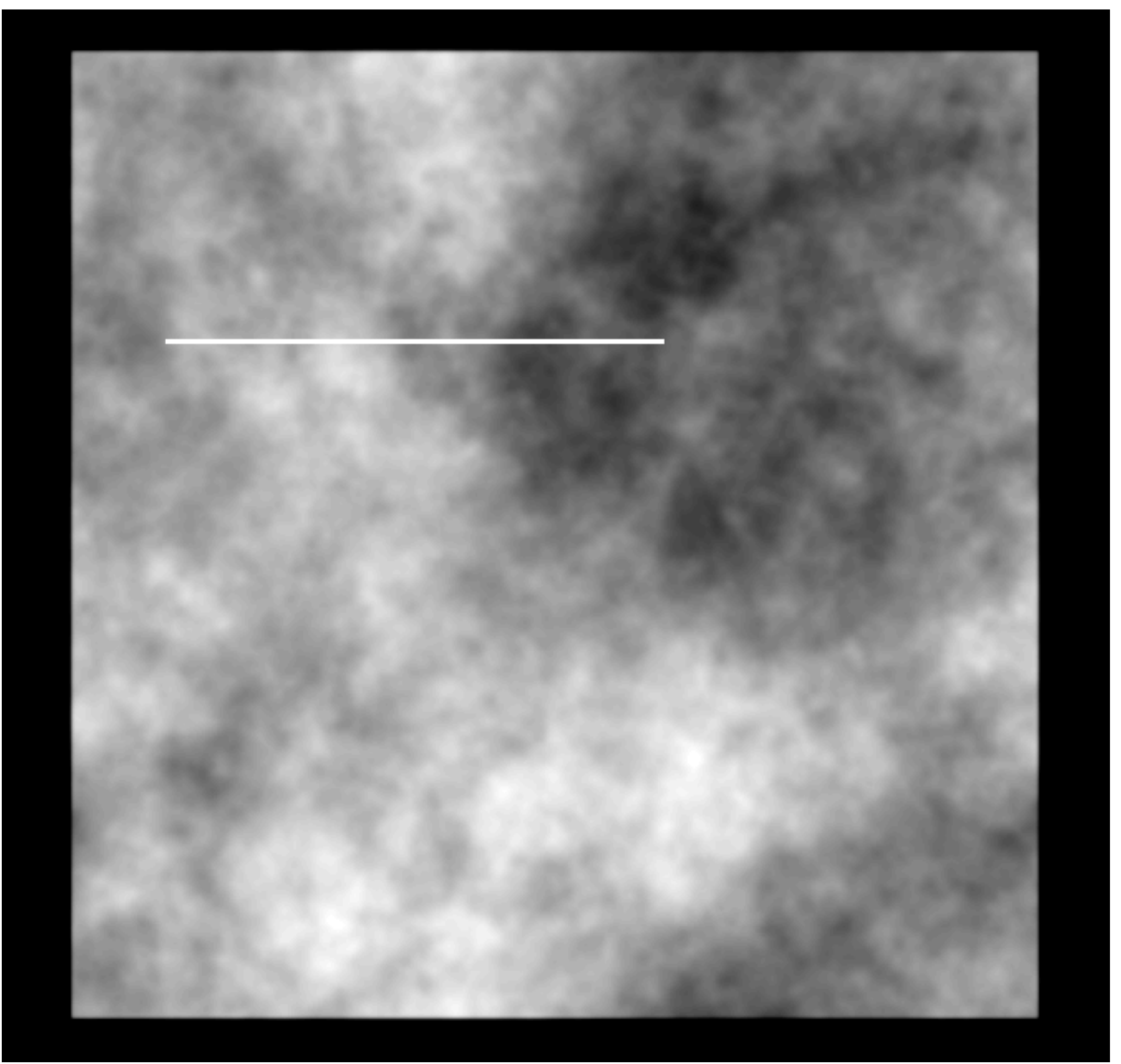}\label{fig:true2}} ~~~~\subfigure[Spectral
  density]
  {\includegraphics[width=0.32\textwidth]{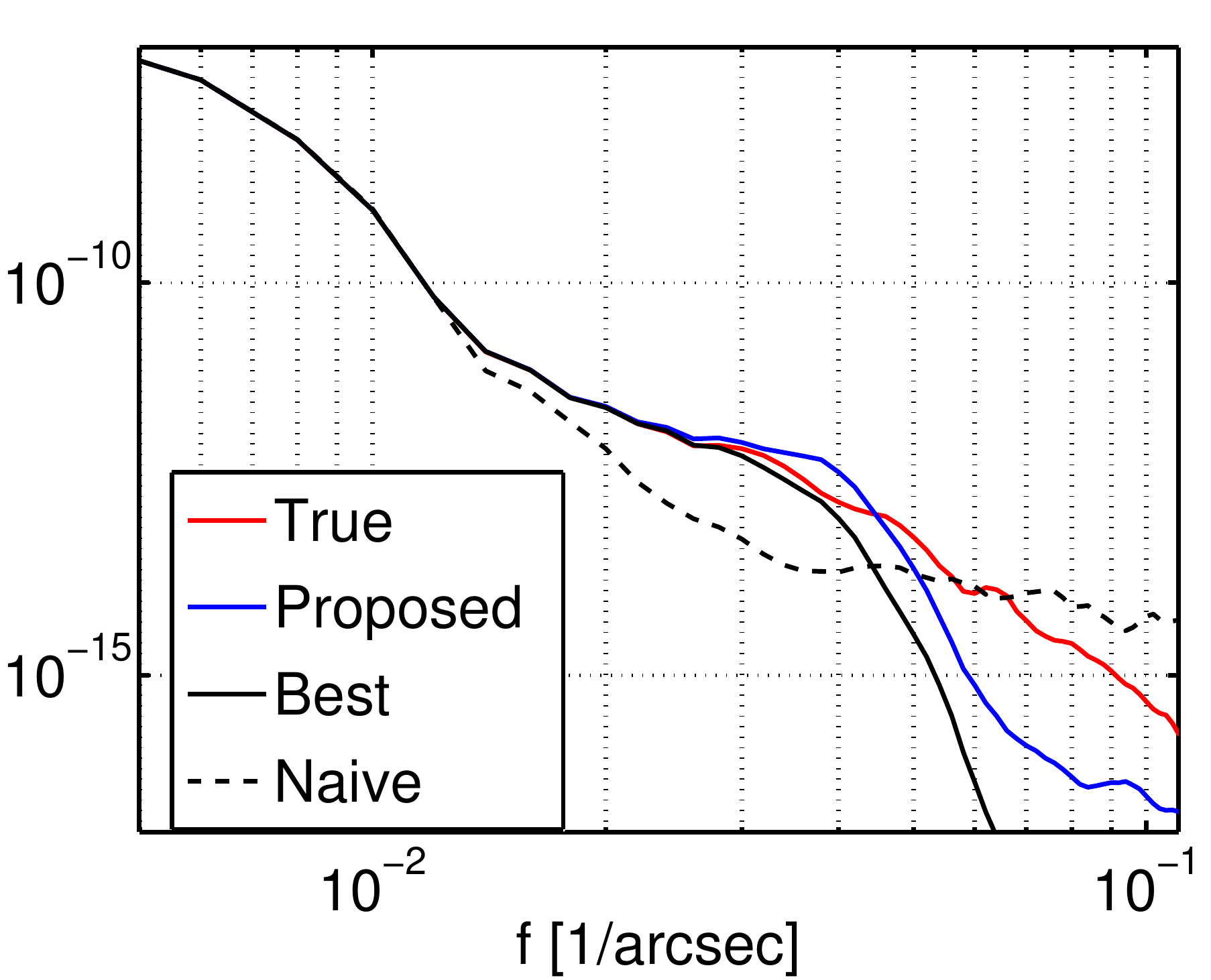}
    \label{fig:psdEapCirrus}}\\

  \caption{Comparison of reconstructed map for the Galactic Cirrus:
    proposed map (Fig.~\ref{fig:eapCirrusIm}), best map
    (Fig.~\ref{fig:cirrusDiffRegNormal}), naive map
    (Fig.~\ref{fig:naiveMap}), and true map (Fig.~\ref{fig:true2}).
    Fig.~\ref{fig:sliceEapCirrus} shows a profile (marked by the white
    line in Fig.~\ref{fig:true2}) and Fig.~\ref{fig:psdEapCirrus} the
    spectrum (circular means of power spectra). Uncertainties are
    given in Fig.~\ref{fig:gibbsCirrus} and quantitative results are
    given in Tab.~\ref{tab:ErreurMap}. Comments are given in
    section~\ref{sec:etude-sur-le}. \label{fig:gibbsCirrus2}}

\end{figure*}

The qualitative and quantitative assessment of the reconstructed maps
is presented here for the Galactic Cirrus; the first results are shown
in Fig.~\ref{fig:gibbsCirrus2}.
\begin{itemize}
\item The unsupervised method (proposed method) is outlined in
  Fig.~\ref{fig:eapCirrusIm}. The hyperparameters are automatically
  set (without knowing of the sky truth).
\item The best-supervised method is outlined in
  Fig.~\ref{fig:cirrusDiffRegNormal}. The hyperparameters are set by
  hand to minimize the error $\Ec$ (knowing the sky truth). It is
  referred to as the best map and was previously presented in our
  paper~\citep{Orieux12}.
\item The naive map (coaddition) and the true map are shown in
  Figs.~\ref{fig:naiveMap} and~\ref{fig:true2}.
\item In addition, Fig.~\ref{fig:sliceEapCirrus} gives spatial
  profiles (vertical in the middle of the map) and
  Fig.~\ref{fig:psdEapCirrus} gives the spectral\footnote{This
    spectrum is computed from the FFT-2D of the map by averaging the
    coefficients in regularly spaced concentric rings. This gives a 1D
    spectrum containing the isotropic approximation of the spectral
    map properties.} profiles.
\end{itemize}
As expected and shown in \citep{Orieux12}, the inversion based on an
accurate instrument model considerably improves the quality of the
map: see the proposed map and the best map compared to the naive map
and to the true map. The proposed map is visually very similar to the
true map. In particular, our method restores details of small spatial
scales (with spectral extension from null to high frequency) that are
invisible on the naive map but are present on the true map (see
Fig.~\ref{fig:sliceEapCirrus}). In addition, our method correctly
restores the structures of large spatial scales (low frequencies) and
also the mean level of the map (null frequency), i.e. the photometry.

To assess the pertinence of estimating $\gB$ and $\gx$ in terms of map
quality, we compared the proposed map with the best map. They are
visually very similar (see Figs.~\ref{fig:eapCirrusIm} and
~\ref{fig:cirrusDiffRegNormal}). In the quantitative terms given by
Tab.~\ref{tab:ErreurMap}, the best map produces an error $\Ec$ of
3.12\% and the proposed map produces an error $\Ec$ of 3.47\%, which
is only slightly higher. In other words, the proposed unsupervised
method automatically (without knowing the sky truth) determines
hyperparameters that produce a map almost as good as the best map
(which requires knowing the sky truth).

\begin{figure*}[htbp]
  \centering

  \subfigure[Map of PSD]{%
    \includegraphics[width=0.25\textwidth]{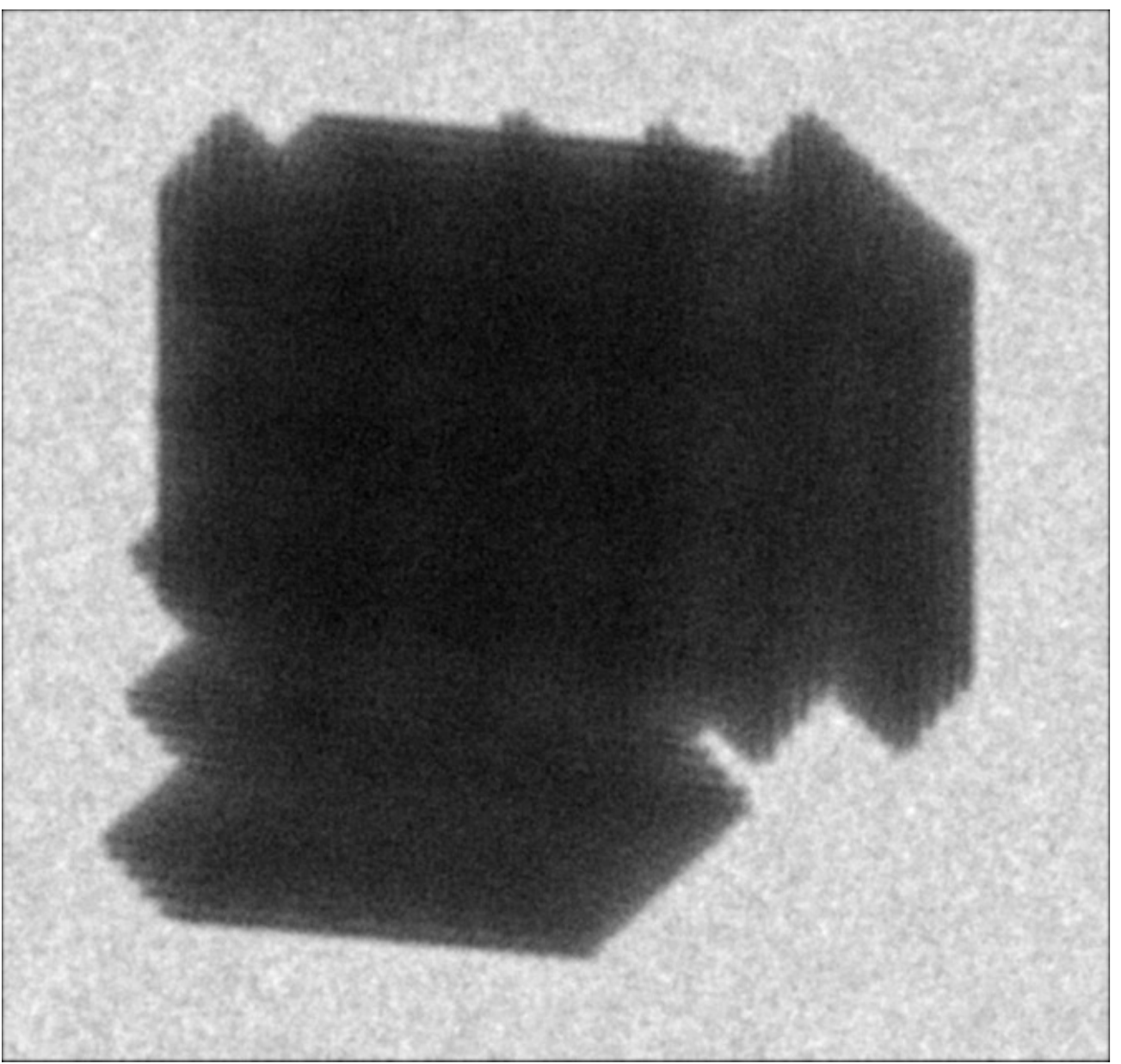}
    \label{fig:stdNScirrus}}%
  ~~~%
  \subfigure[True map and PM~$\pm$~PSD]{%
    \includegraphics[width=0.3\textwidth]{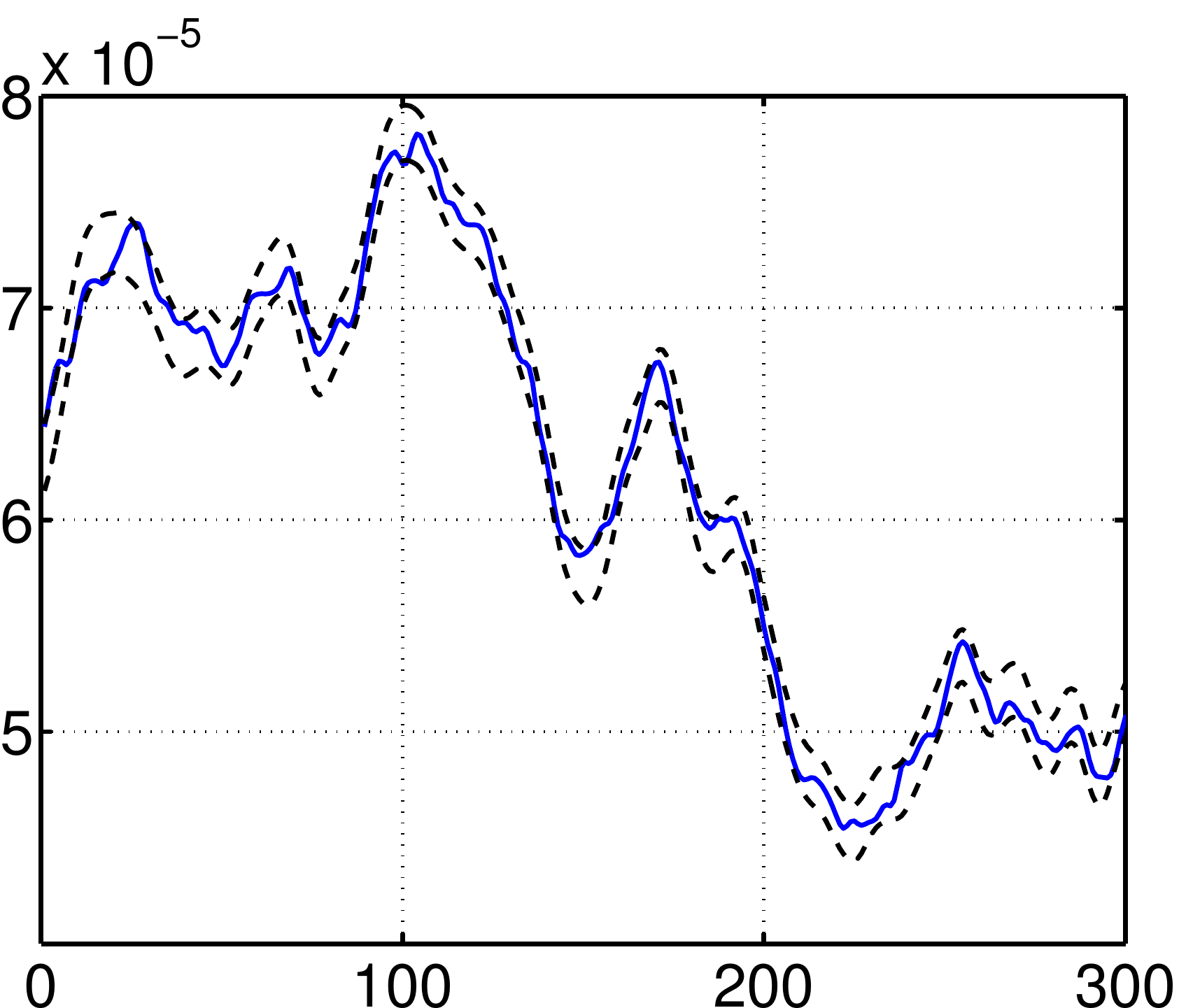}
    \label{fig:slicestdNScirrusPM}}%
  ~~~%
  \subfigure[True spectrum and PM~$\pm$~PSD]{%
    \includegraphics[width=0.3\textwidth]{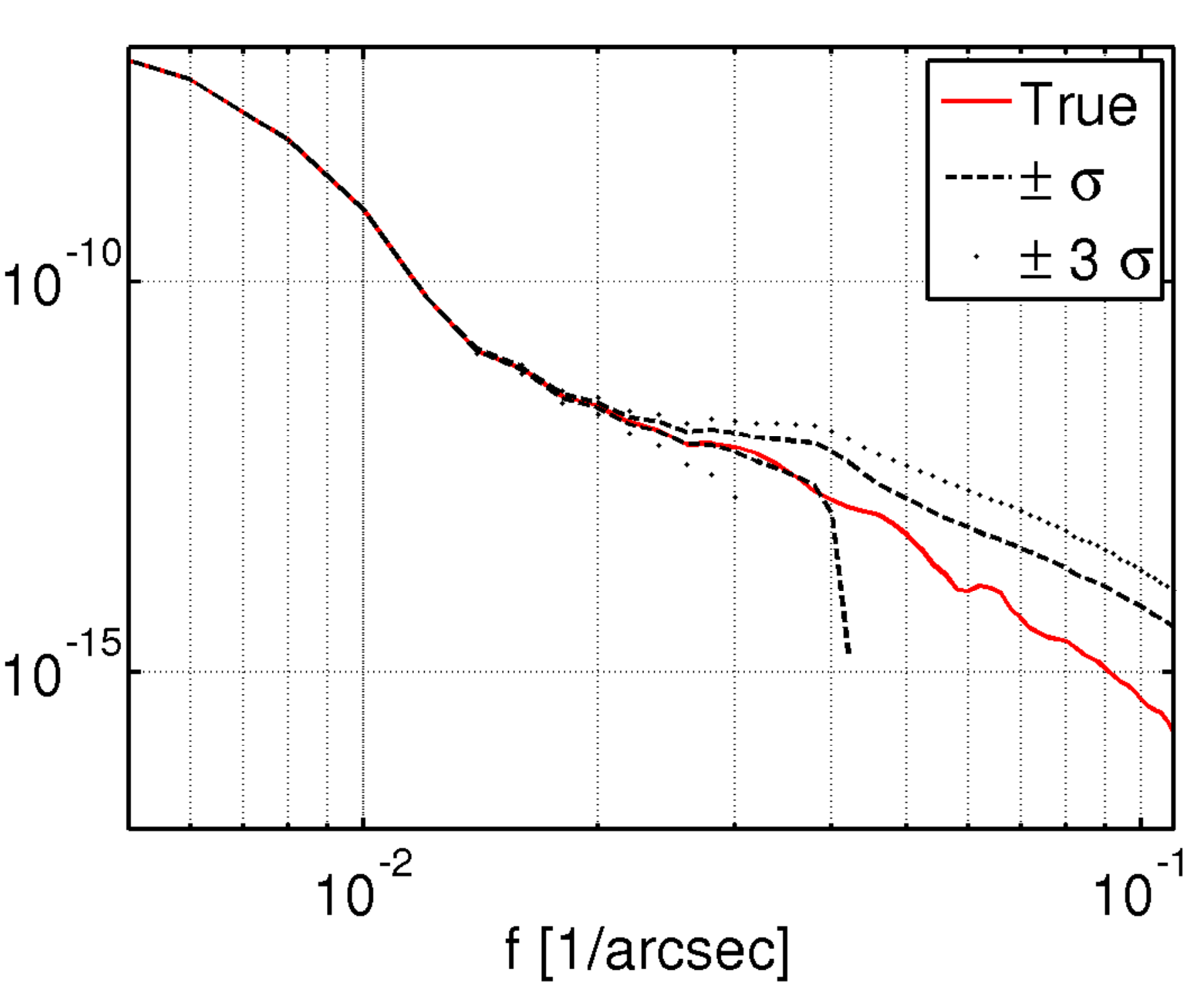}
    \label{fig:psdstdNScirrusPM}} ~~~

  \caption{PSD and quantification of
    uncertainties. Fig.~\ref{fig:stdNScirrus} shows the map of the PSD
    and Figs.~\ref{fig:slicestdNScirrusPM}
    and~\ref{fig:psdstdNScirrusPM} show the interval around the
    \textit{estimated map} $\pm$ PSD and the true map. In the spatial
    domain, Fig.~\ref{fig:slicestdNScirrusPM} is a profile (marked by
    the white line on~\ref{fig:true2}) and in the spectral domain
    Fig.~\ref{fig:psdstdNScirrusPM} is the circular means of the power
    spectra. \label{fig:gibbsCirrus}}
\end{figure*}

\begin{figure*}[htpb]
  \centering

  \subfigure[Proposed map residuals]{%
    \includegraphics[width=0.3\textwidth]{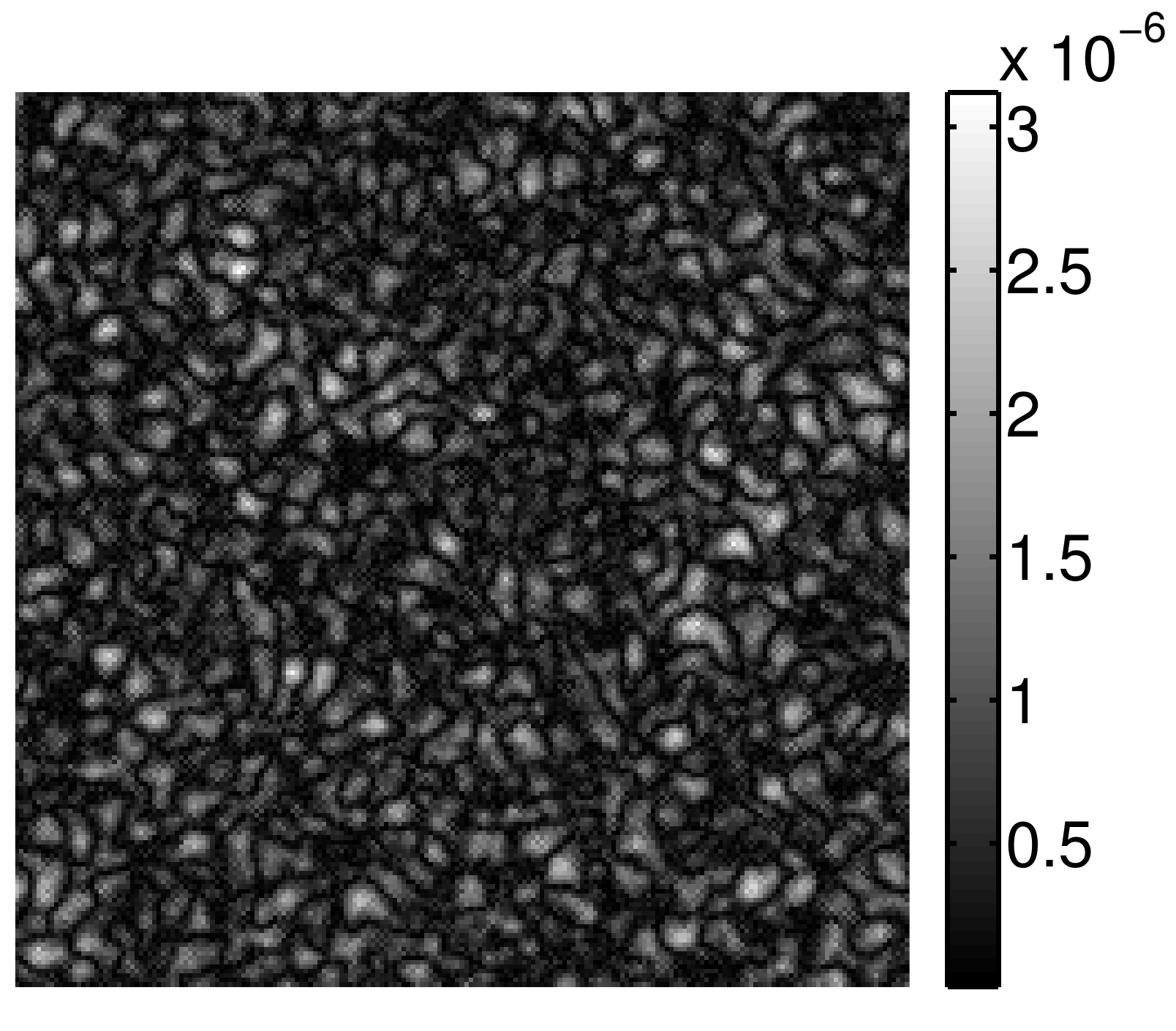}
    \label{fig:coadd_residal}}%
  ~~~%
  \subfigure[Best map residuals]{%
    \includegraphics[width=0.3\textwidth]{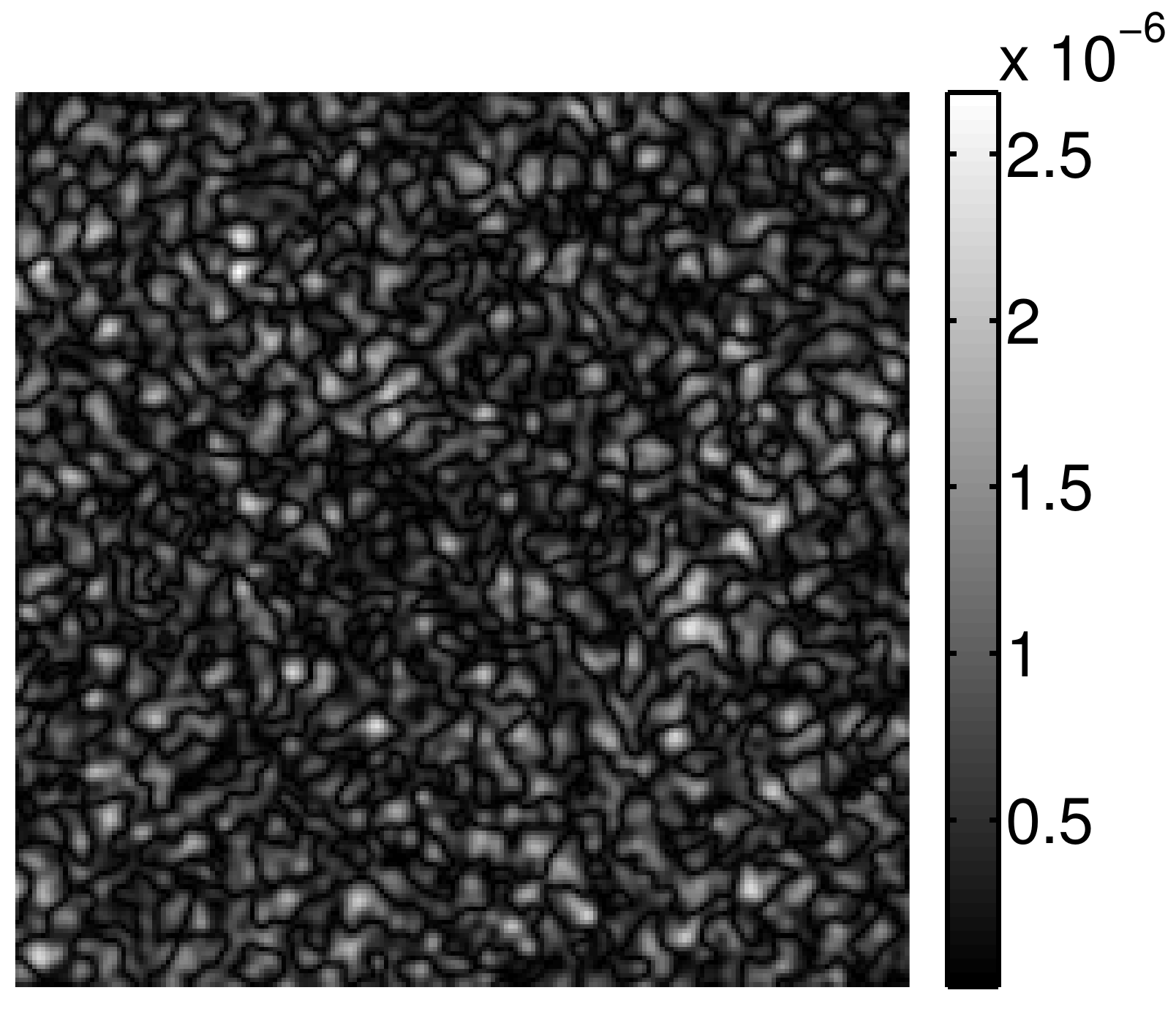}
    \label{fig:optim_residal}}%
  ~~~%
  \subfigure[Naive map residuals]{%
    \includegraphics[width=0.3\textwidth]{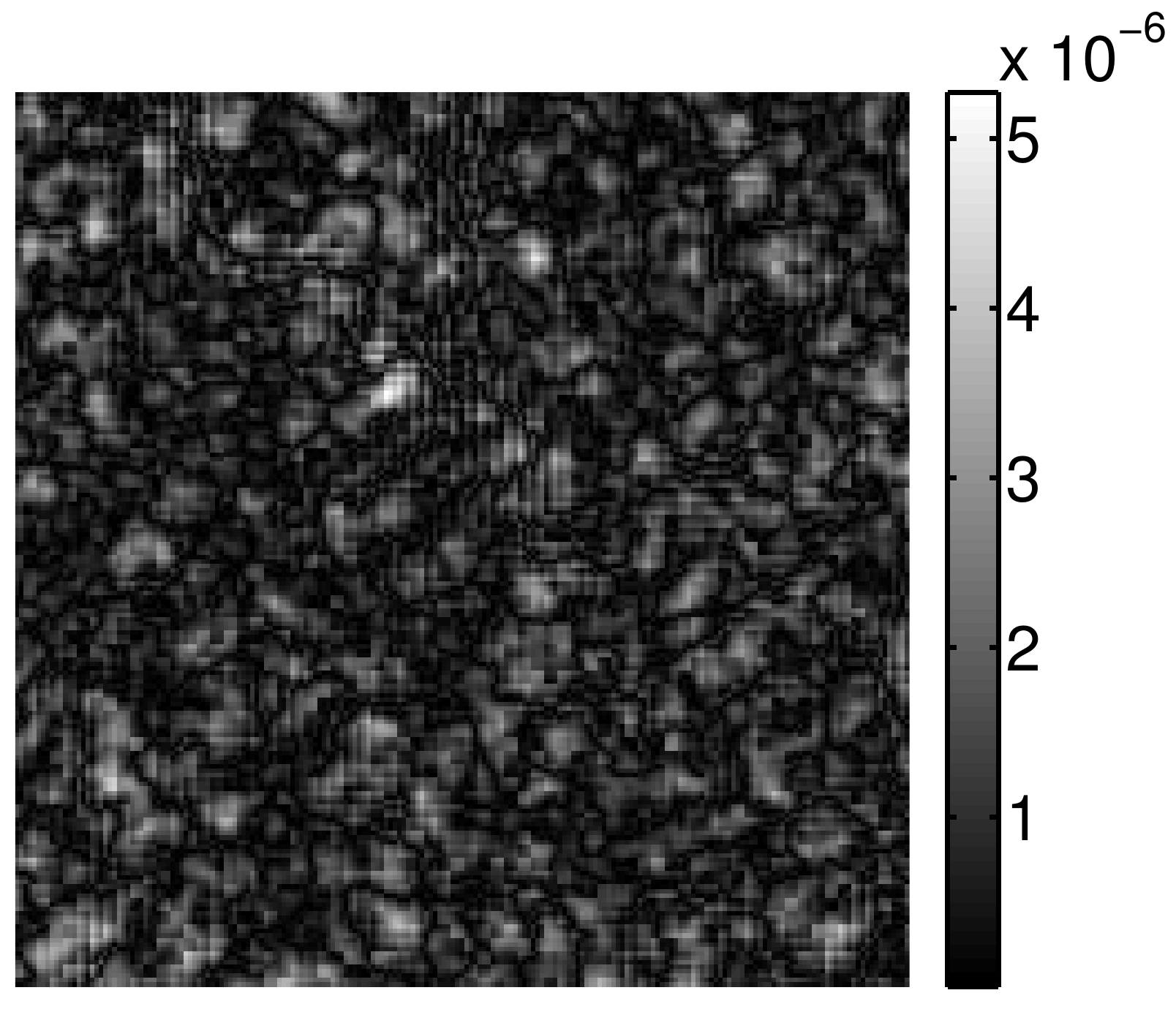}
    \label{fig:coadd_residal}}%
  ~~~%

  \caption{Residuals of maps in the central part of measured
    coefficiens. This illustrates the grainy structure of the proposed
    map wrt. best map. The naive map residuals suffer twice more
    errors and a squared feature caused by the pixel model.}

  \label{fig:residuals}
\end{figure*}

\begin{table}[htbp]
  \centering
  \begin{tabular}{c|c}
    & Reconstruction error $\Ec$ \\
    \hline
    Unsupervised ($\widehat\gx$)       & 0.016 \% \\
    \hline
    Best supervised ($\gx^{\rm best}$) & 0.0129 \% \\
    \hline
    Naive map                          & 0.0435 \% \\
    \hline
  \end{tabular}
  \bigskip
  \caption{Comparison of reconstruction error $\Ec$ (see
    Eq.~\eqref{Eq:DefErreurL2}) for the Galactic Cirrus (the error $\Ec$
    only accounts for the observed area of the map). The proposed approach
    (which does not require knowing the true map) produces an error
    only very slightly higher than the best map (which does require
    knowing true map).\label{tab:ErreurMap}}
\end{table}

However, the proposed map shows a fine grainy texture that is visible
neither in the true nor in the best map. This feature is also visible
on the residual map of the coefficiens
(Fig.~\ref{fig:residuals}). This is also observable in
Fig.~\ref{fig:psdEapCirrus}: the spectrum of the proposed map passes
above the spectrum of the true map in the spectral band
$0.025$--$0.035\,\mathrm{arcsecond}^{-1}$. This defect is related to a
slight overevaluation of the observation contribution with respect to
the \prior contribution. It is referred to as under-regularization and
yields an overamplification of the observation in this spectral
band. This confirms the behaviour previously observed in
deconvolution~\citep{Orieux10a,Giovannelli08} or noted for the maximum
likelihood~\citep{Fortier93}. Nevertheless, it is remarkable that this
defect is correctly notified by the PSD, as explained in the next
paragraph.

Indeed, the approach naturally provides a measure of reliability
through the PSD shown in Fig.~\ref{fig:gibbsCirrus}. Two zones can be
seen in Fig.~\ref{fig:stdNScirrus}, in accordance with
Fig.~\ref{fig:balay}: the central zone (where observations are
available) and the peripheral zone (extrapolated from observations of
the central zone and based on the \prior regularity). The boundary
between the two zones also exhibits the variation of the observation
hit and scanning strategy notably well. In addition, the posterior
standard deviation also illustrates the difference between the zones
observed with our without cross scan. From a spatial standpoint,
Fig.~\ref{fig:slicestdNScirrusPM} shows an interval around the
\textit{estimated map} with plus/minus PSD. The main result is that
the \textit{true map} is clearly within the interval. In a similar
way, from a spectral standpoint the results are given by
Fig.~\ref{fig:psdstdNScirrusPM} (in relation with
Fig.~\ref{fig:psdEapCirrus}): the \textit{true spectrum} is also
within the interval. More specifically, incorrectly reconstructed in
the spectral band (above $0.025\,\mathrm{arcsecond}^{-1}$), the
stronger PSD \emph{clearly shows} that the estimated spectrum is
certainly submitted to unsatisfactory errors or confidence.

\subsubsection{Assessment of hyperparameter estimation}
\label{sec:exp2}\label{sec:chaines-des-hyper}

\begin{figure}[htb]
  \centering

  \subfigure[$\gB$ chain]{\includegraphics[width=0.22\textwidth]%
    {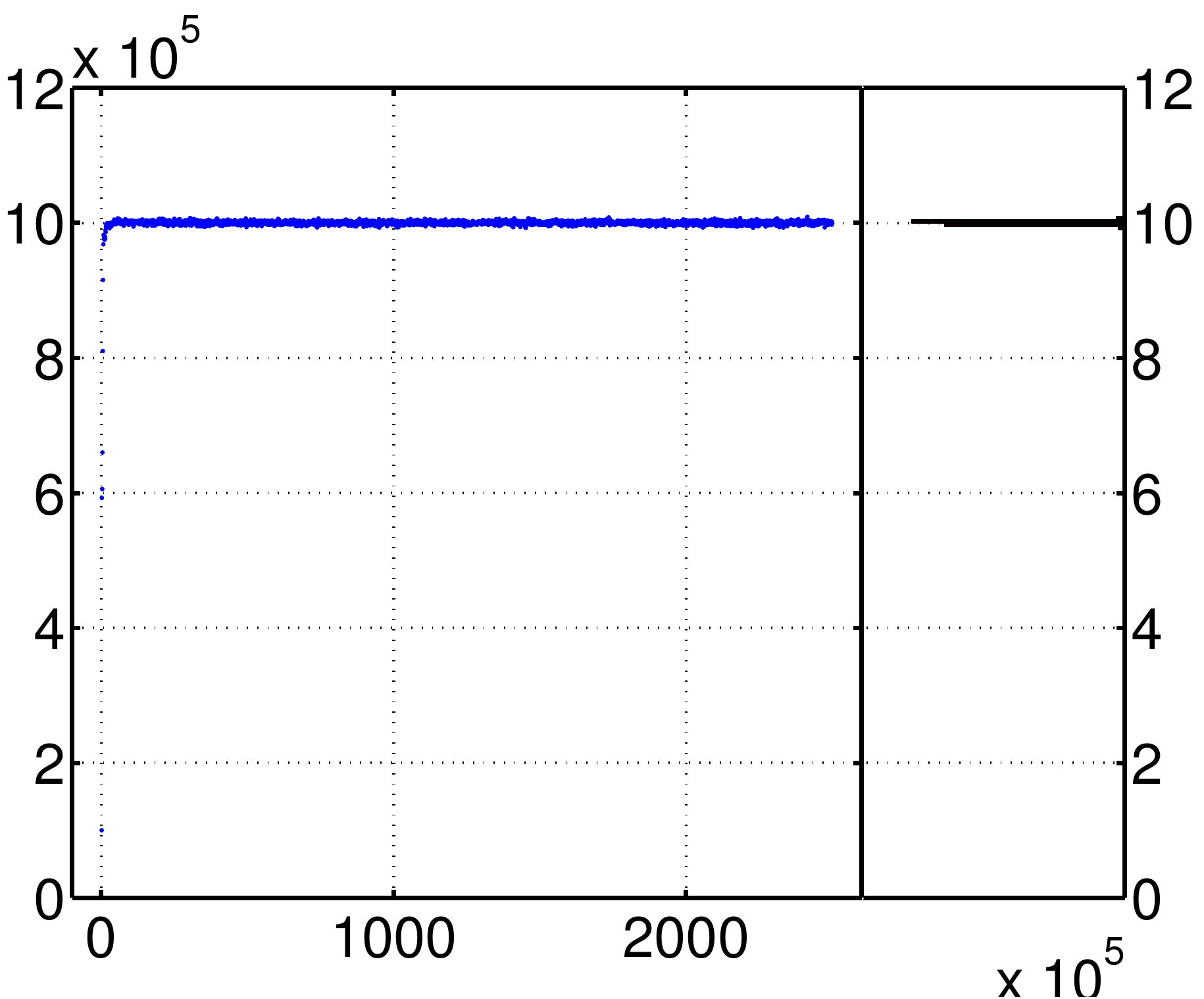} \label{fig:cgB}}~~~~~~%
  \subfigure[$\gx$ chain]{\includegraphics[width=0.22\textwidth]%
    {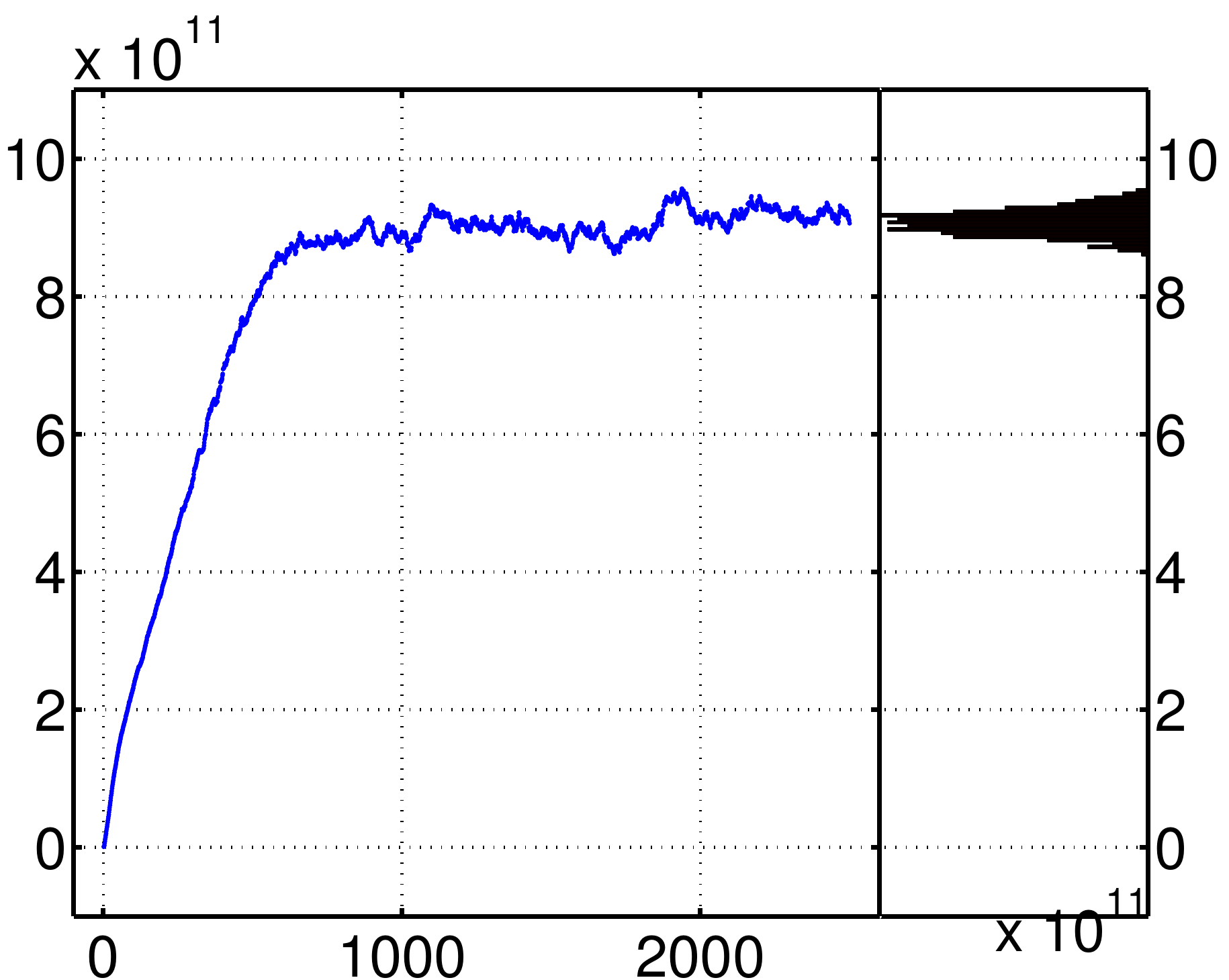} \label{fig:cgX}}

  \caption{Chains and histograms for $\gB$, Fig.~\ref{fig:cgB}, and
    $\gx$, Fig.~\ref{fig:cgX}, for the Galactic Cirrus. The chains
    show the burn-in period (about $1000$ iterations) and the steady
    state. The corresponding histograms are computed on steady state
    only. \label{fig:hyperHisto}}

\end{figure}

This section assesses the unsupervised capabilities through evaluating
the hyperparameter estimation using the Galactic Cirrus and a
realization of the \prior for the map (which makes a true value
$\gx^*$ available). Fig.~\ref{fig:hyperHisto} shows the chains and the
histograms that approximate the marginal \post densities $p(\gB|\yb)$
and $p(\gx|\yb)$. In both cases, the histogram is relatively narrow
although the \prior is a wide non-informative Jeffreys' distribution
(see Eq.~\eqref{Eq:PriorJeffreys}). In other words, the observations
are sufficiently informative to quantify noise and regularity level
and the method is able to capture this information.

From a quantitative standpoint, results are given in
Tab.~\ref{tab:estimHypers}. For the Galactic Cirrus and \prior
realization, the estimated values $\widehat \gB$ are very similar to
the true value $\gB^{*}$ (error is less than 1\,\%). Moreover, the PSD
are very low (0.40\,\%). In the case of \prior realization, the
estimated value $\widehat\gx$ is in the correct range but the error is
larger (about 17\,\%) and the PSD is 1.7\,\%. This difference can be
naturally explained by two elements: (i) the noise is added at the
system output, so it is directly observed, whereas the map is at the
system input, i.e. indirectly observed; (ii) the added noise is a
realization of the \prior density for the noise while the Galactic
Cirrus is not a realization of the \prior density for the map.

\begin{table*}[htbp]
  \centering
  \begin{tabular}{c|ccc|ccc|c}
    & $\gB^{*}$ &    $\widehat\gB$   &        PSD & $\gx^{*}$     &     $\widehat\gx$    &         PSD         & $\gx^{\rm best}$\\\hline
    Cirrus  &  $10^6$   & $1.009\times 10^6$ & $4.07\times 10^3$ & - & \numprint{8.99e11}  & \numprint{2.46e10}  & \numprint{2.47e12} \\
    Prior   &  $10^6$   & $1.003\times 10^6$ & $4.05\times 10^3$ & $4\times 10^{11}$& $3.28\times 10^{11}$ & $1.07\times 10^{10}$& $8.37\times 10^{11}$ \\
  \end{tabular}
  \caption{Hyperparameter estimation: true values, estimates, PSD and
    best values.\label{tab:estimHypers}}
\end{table*}

\subsubsection{Real observation processing}
\label{sec:hyperp-estim-real}

This section proposes a first assessment for a real observation. It is
based on the reflection nebula NGC\,7023 acquired during the science
demonstration phase of Herschel, which as been presented in
\citep{Abergel10} and was processed in our previous
paper~\citep{Orieux12}. There and here, computations are made on the
level-1 files processed using HIPE. In our previous paper
\citep{Orieux12}, the offsets were removed in a pre-processing step
and the regularization parameter was tuned by hand compromise between
gain in resolution and overamplification of the observations. In
contrast, here both are automatically tuned.

\begin{figure*}[htbp]
  \centering

  \subfigure[$\gB$ (noise parameter) \label{fig:hyper-real-data-gnchain}]{%
    \includegraphics[width=0.30\textwidth]{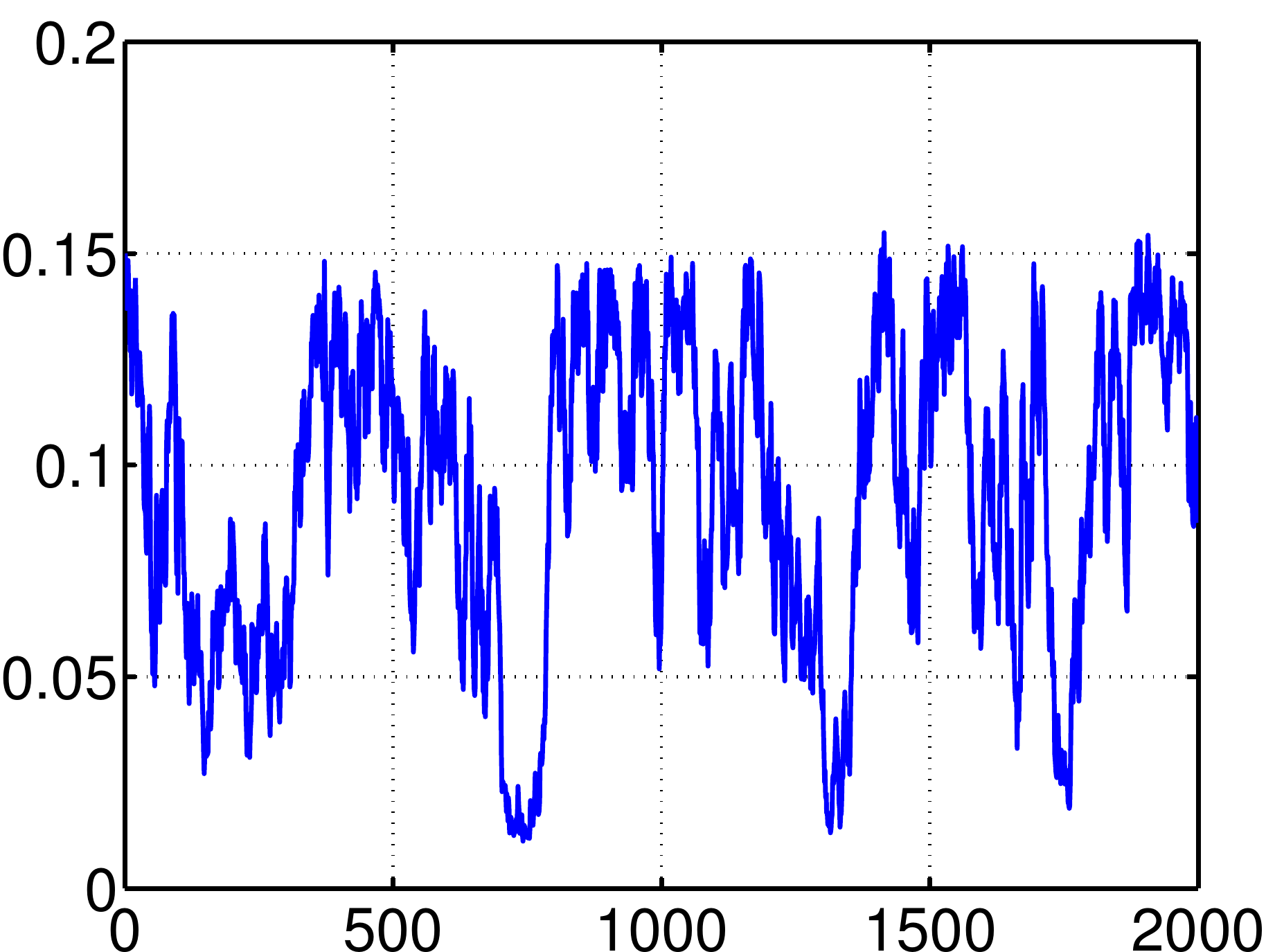}}~~~~~%
  \subfigure[$\gx$ (image parameter) \label{fig:hyper-real-data-gxchain}]{%
    \includegraphics[width=0.28\textwidth]{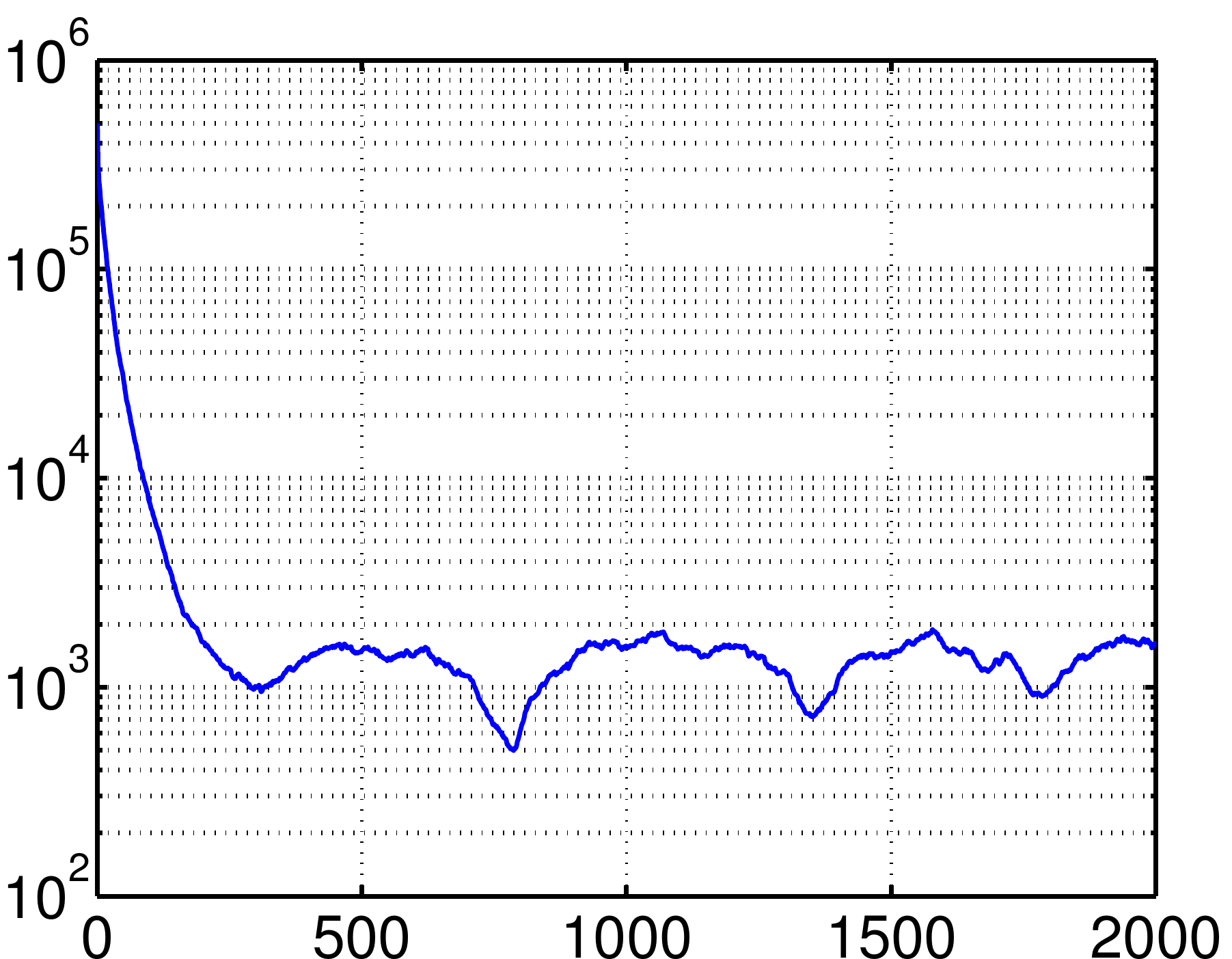}}~~~~~%
  \subfigure[Reconstructed map\label{fig:hyper-real-data-image}]{%
    \includegraphics[width=0.30\textwidth]{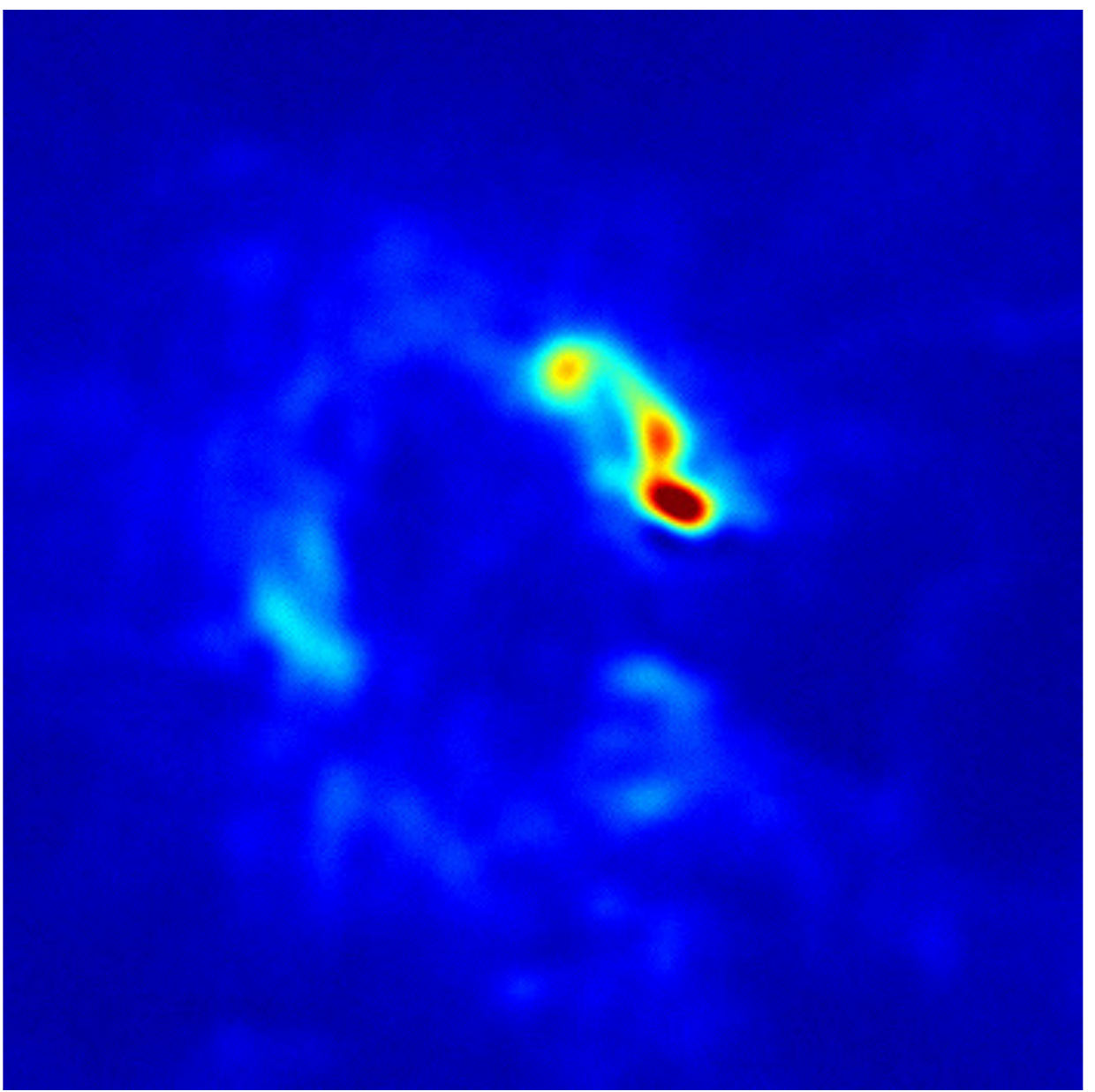}}

  \subfigure[Offset chain for the first bolometer\label{fig:offset2}]{%
    \includegraphics[width=0.30\textwidth]{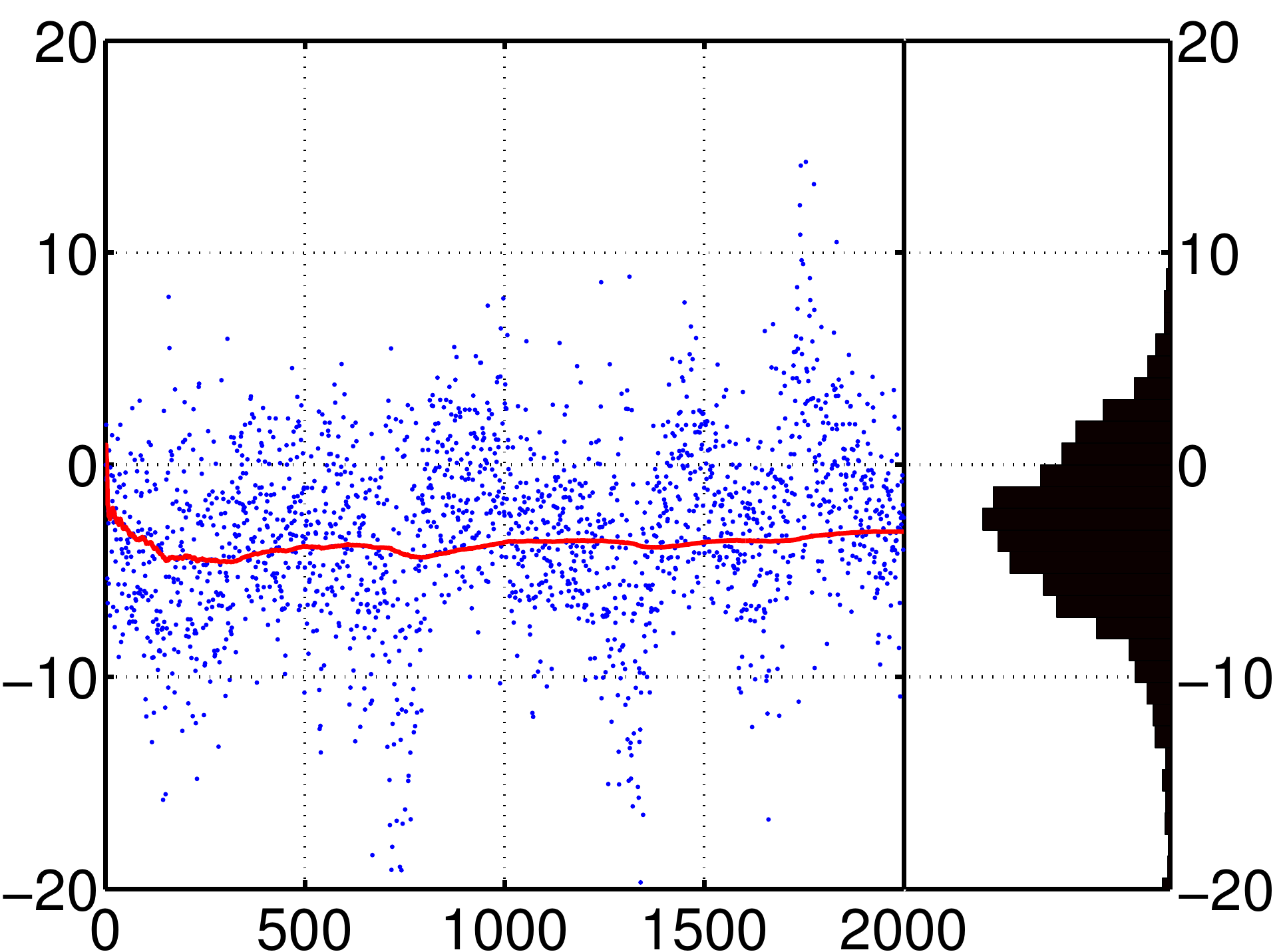}}~~~~~%
  \subfigure[Offset chain for the second bolometer\label{fig:offset54}]{%
    \includegraphics[width=0.30\textwidth]{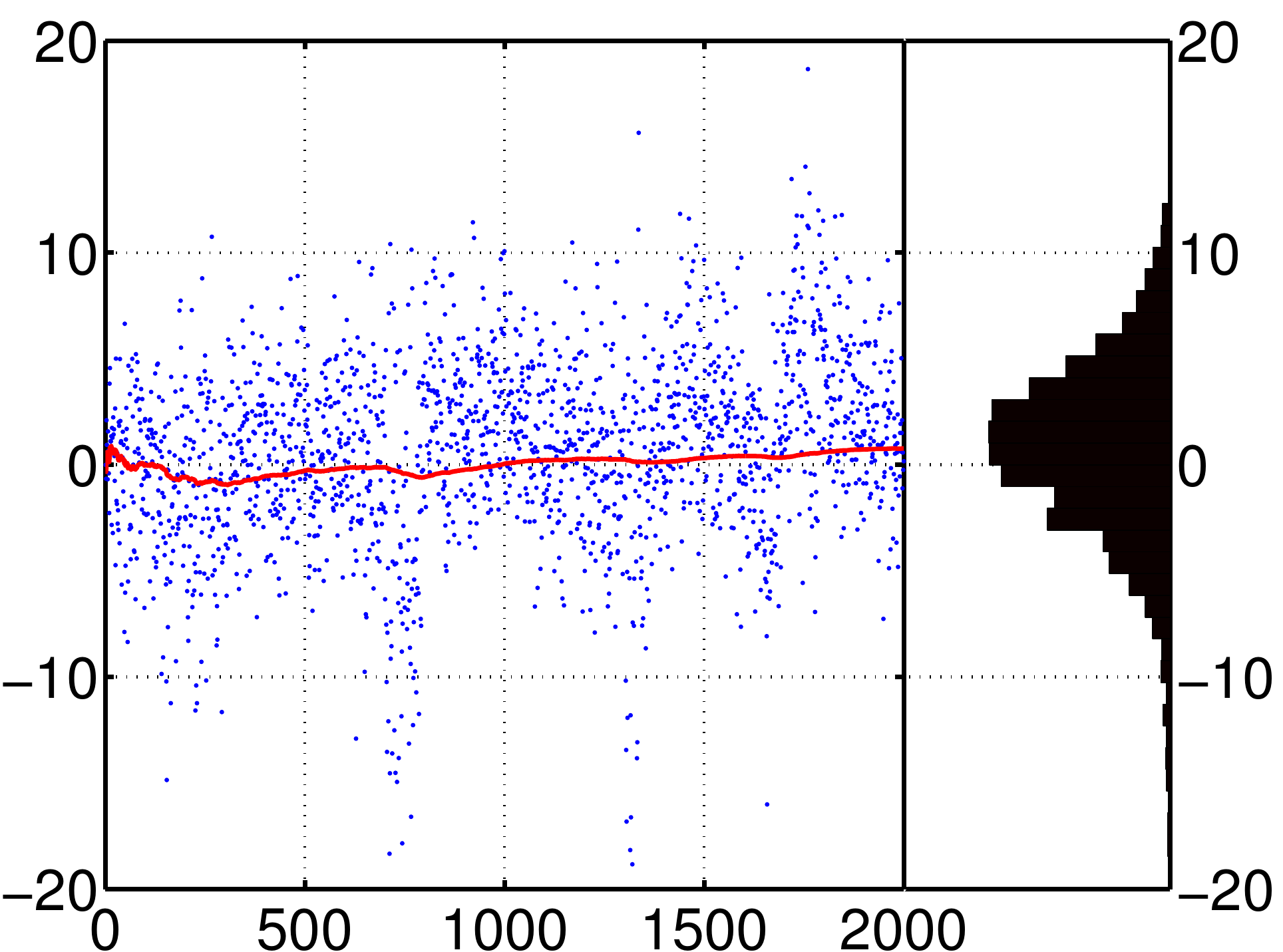}}~~~~~%
  \subfigure[Naive map (coaddition)\label{fig:hyper-real-data-image-coadd}]{%
    \includegraphics[width=0.30\textwidth]{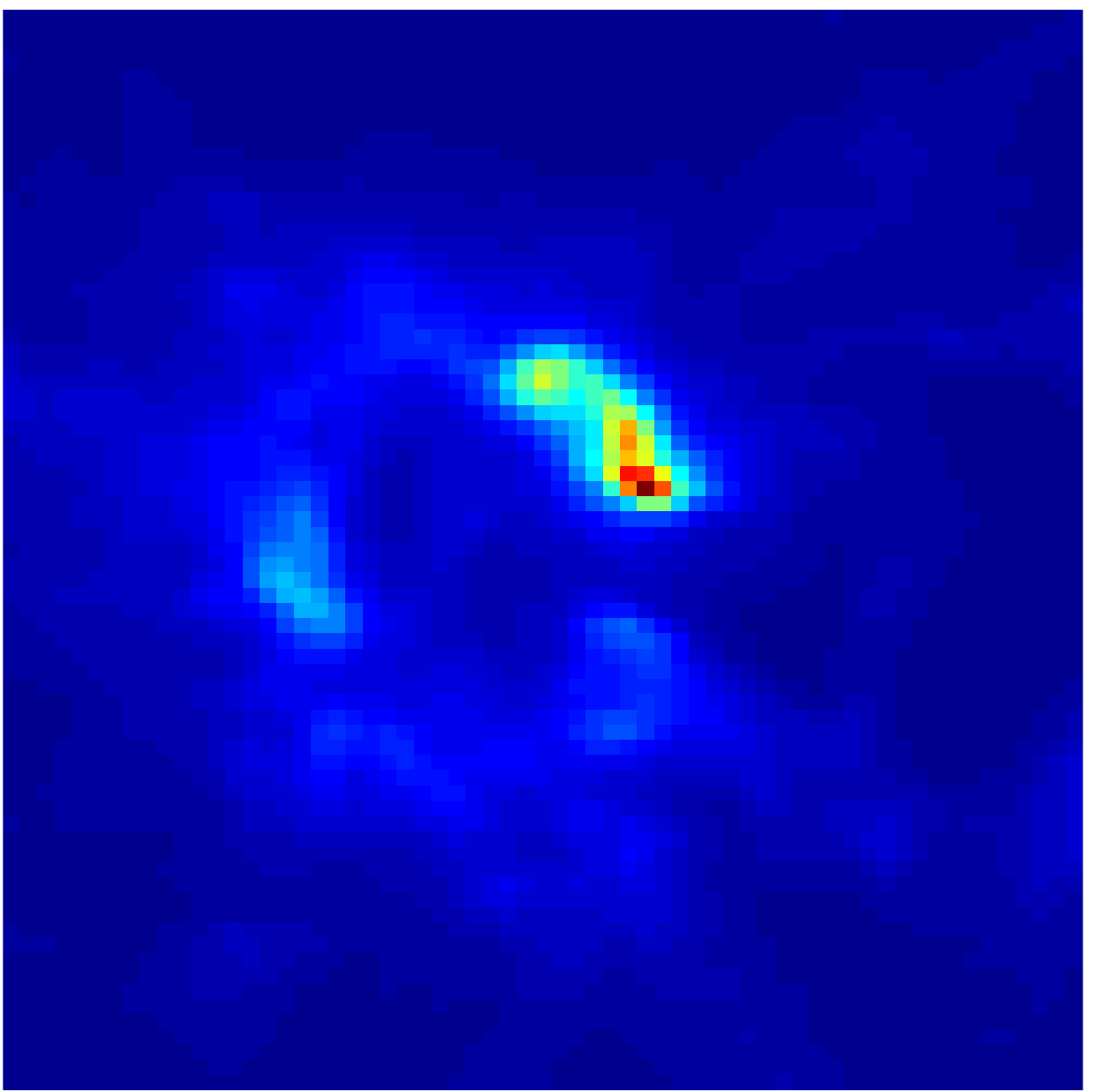}}

  \caption{Results for real observation processing (reflection nebula
    NGC\,7023). Chains for the noise parameter ($\gB$) and for the
    image parameter ($\gx$) in Figs.~\ref{fig:hyper-real-data-gnchain}
    and~\ref{fig:hyper-real-data-gxchain}. The stationary state is
    attained after a burn-in time of about $500$
    iterations. Fig.~\ref{fig:hyper-real-data-image} shows the
    corresponding map. Figs.~\ref{fig:offset2} and \ref{fig:offset54}
    illustrate chains and marginal histogram for two bolometer
    offsets. \label{fig:hyper-real-data}}
\end{figure*}

Fig.~\ref{fig:hyper-real-data} presents the evolution of the chains
for the hyperparameters: Fig.~\ref{fig:hyper-real-data-gnchain} for
the noise parameter $\gB$ and Fig.~\ref{fig:hyper-real-data-gxchain}
for the image parameter $\gx$. It is important to notice that the
algorithm behaves in a very similar manner for the real observation
and for the simulated observation (see Fig.~\ref{fig:hyper-real-data}
compared to Fig.~\ref{fig:hyperHisto}). The figures also give an
empirical indication of the algorithm operation: after a burn-in time
(empirically less than about $500$ iterations) the stationary state is
attained and the chain remains in a steady state: the samples are
drawn under the posterior density. Concerning the offsets, the chains
begin in the steady state, thanks to a good initialization based on
\citep{Orieux12} results. All bolometer offsets behave in the same
manner with two example illustrated Figs.~\ref{fig:offset2} and
\ref{fig:offset54}. The empircal mean of the offsets sample start to
stabilize at approximately 500 samples.

Fig.~\ref{fig:hyper-real-data-image} shows the corresponding
reconstructed map. Its quality is equivalent to the quality of the map
restored by empirically tuning the hyperparameter presented in
\citep{Orieux12}, Fig.~8. In other words, the proposed unsupervised
method automatically determines hyperparameters (noise power and
offsets as well as sky power) that produce a map almost as good as the
map produced by a hand-made hypermarameter tuning. In addition, the
map remains far better than the naive map shown in
Fig.~\ref{fig:hyper-real-data-image-coadd}.

\subsection{Myopic and unsupervised approach}
\label{sec:instr-param-estim}

The myopic and unsupervised question is a threefold problem that is
much more ambitious: estimate the instrument parameter, the
hyperparameters, and the map itself from a unique observation. In
addition, the instrument parameter intervenes in a complex way in the
description of the observations, and moreover, the problem is stated
in a context that is doubly delicate: ill-posedness and
high-resolution.

In \citep{Orieux12}, the equivalent PSF has a Gaussian shape whose
standard deviation is proportional to the wavelength: $\Smc(\lambda) =
\eta \lambda$. It is then integrated \wrt the wavelength (to include
the spectral extend) and \wrt the time parameter (to account for the
bolometer response) to form the global instrument response. To test
the method, we consider the instrument parameter $\eta$ to be poorly
known and introduce elements of the feasibility to estimate it.

\begin{figure}[hbtp]
  \centering

  \subfigure[Case~1, parameter chain\label{fig:MyopicPicsChainI}]{%
    \includegraphics[width=0.22\textwidth]{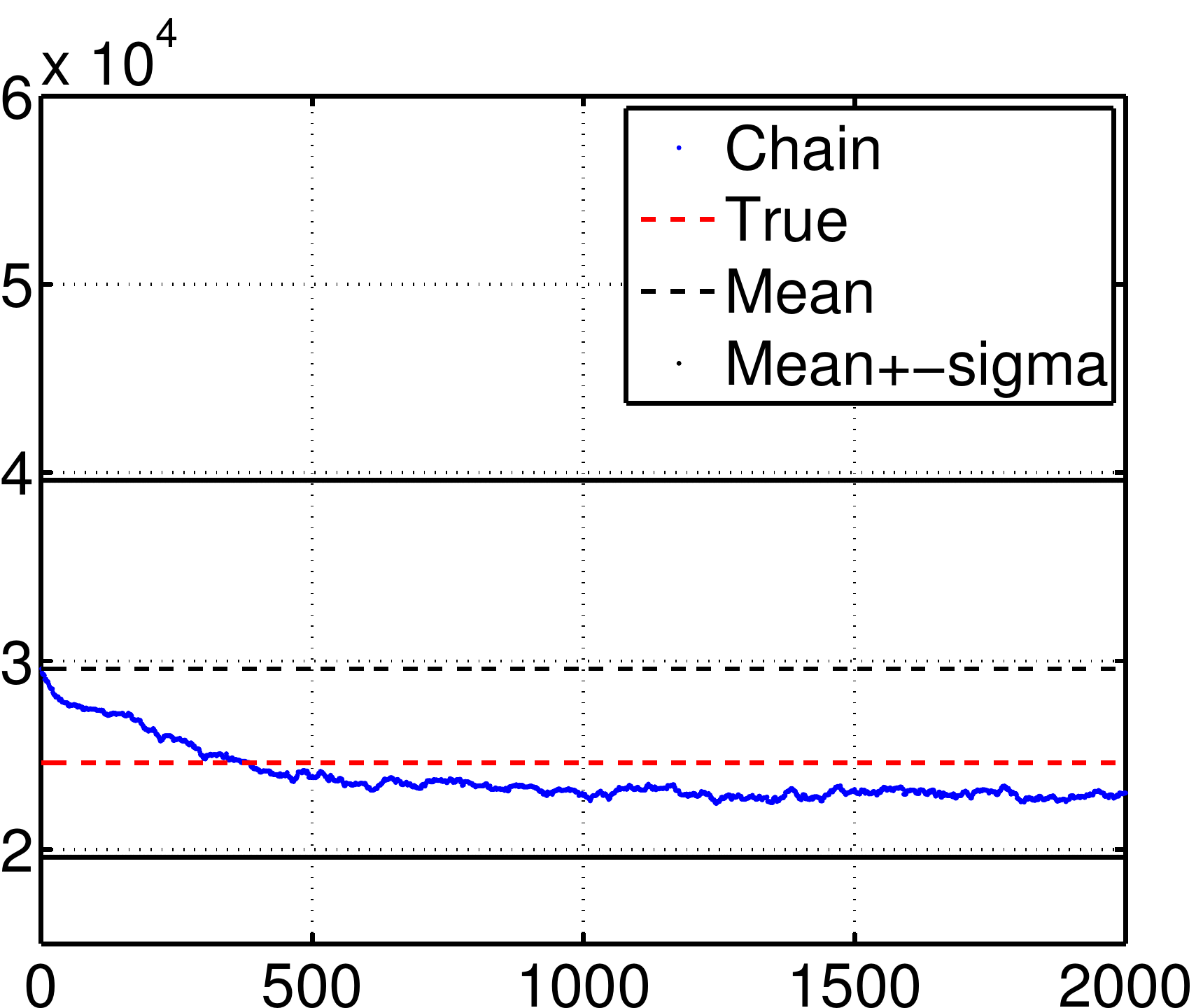}}~~~~~%
  \subfigure[Case~2, parameter chain\label{fig:MyopicPicsChainII}]{
    \includegraphics[width=0.22\textwidth]{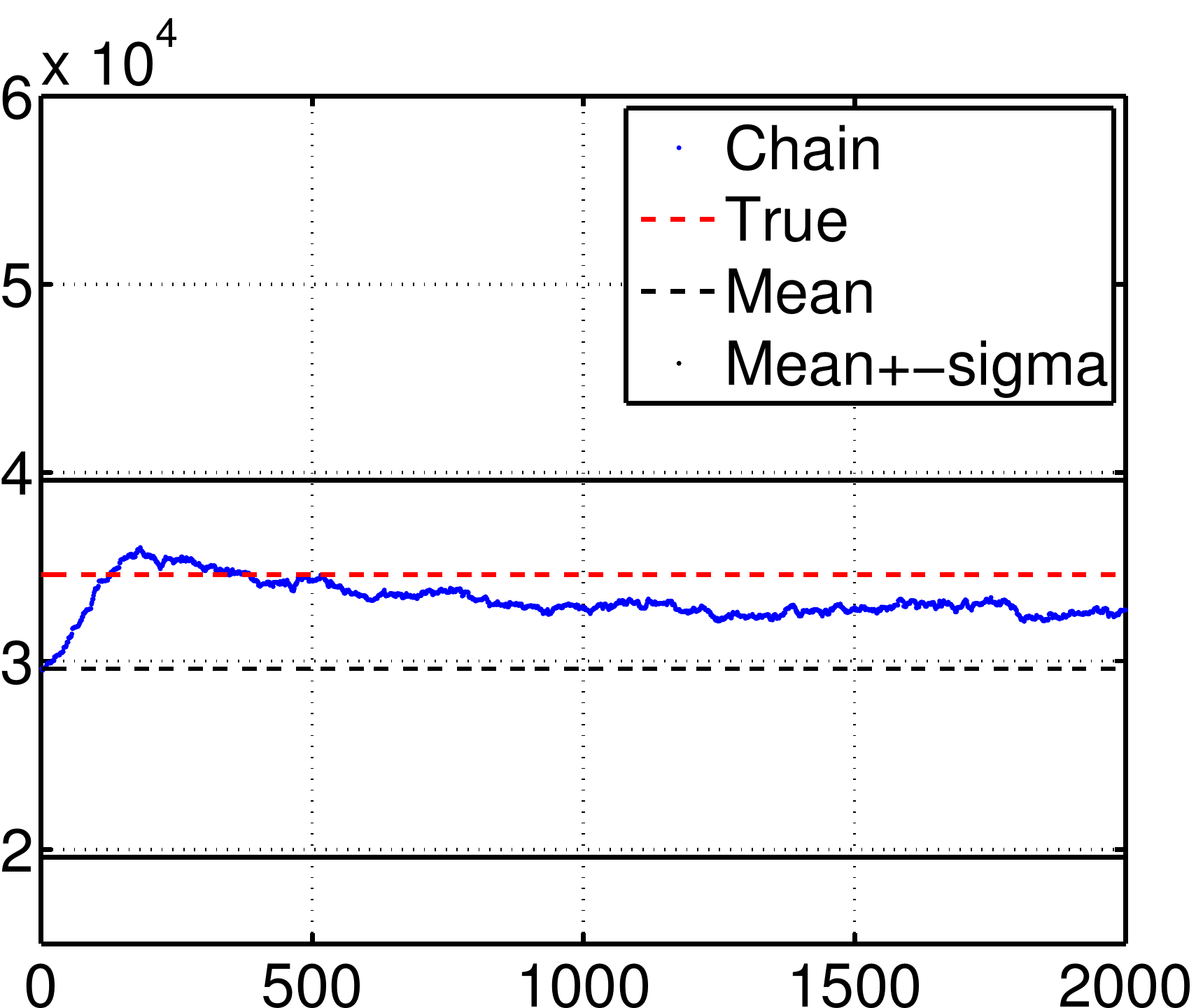}}

  \caption{Instrument parameter chain (myopic and unsupervised
    approach) for the of Galactic Cirrus with point sources. Left
    (right) part of the figure deals with case~1,
    i.e. $\eta^{*}_1=2.46\times 10^4$ (case~2,
    i.e. $\eta^{*}_2=3.46\times 10^4$). The horizontal axis gives the
    iteration index and the vertical range is the prior interval in a
    two-standard-deviations sense. The true value is shown by the
    straight line.  \label{fig:myopicpics}}
\end{figure}

The \prior is the Gaussian density given by
Eq.~\eqref{Eq:PriorInstru}, with $K=1$. Its mean is taken from the
SPIRE observer manual $\mu = 2.96\times 10^4$ [\arcsec/m] and its
standard deviation is set to $\rho = 10^4$, i.e. a relatively large
uncertainty. It is about $33\%$ of the mean and an equivalent prior
interval is $[0.96\times 10^4 , 4.96\times 10^4]$ in a
two-standard-deviations sense. Two cases are investigated for the true
value (used to simulated observations) : $\eta^{*}_1=2.46\times 10^4$
and $\eta^{*}_2=3.46\times 10^4$.  The conditional \post for $\eta$
(section~\ref{Sec:InstruSample}) does not have a standard form and its
sampling (step~(3) of Tab.~\ref{Algo:Gibbs}) relies on a
Metropolis-Hastings sampler, itself based on a random-walk with a
Gaussian excursion. The size of the excursion was chosen so that the
acceptation rate is around~$50\%$. Two maps are used for the
observation: the Galactic Cirrus and the Galactic Cirrus with point
sources. In each case, the algorithm was run several times from
identical and different initializations, and shows similar qualitative
and quantitative behaviours as those in Fig.~\ref{fig:myopicpics}.

\begin{table*}[htb]
  \centering
  \begin{tabular}{|c|c|c|c|c|c|}
    Case &    $\eta^{*}$     &    $\hat\eta$     & $\hat\eta-\eta^{*}$ & $(\hat\eta-\eta^{*})/\eta^{*}$ & $\bar\sigma$ \\
    \hline
    1   & \numprint{2.46e4} & \numprint{2.29e4} & \numprint{-1.65e3}  & \numprint{6.7} \%        & \numprint{2.2e2} \\
    2   & \numprint{3.46e4} & \numprint{3.27e4} & \numprint{-1.86e3}  & \numprint{5.4} \%        & \numprint{2.9e2} \\
  \end{tabular}
  \caption{Quantitative evaluation of the estimation of the instrument
    parameter using the Galactic Cirrus with point sources. Prior mean and
    standard deviation are $\mu = 2.96\times 10^4$ and $\rho =
    10^4$. \label{tab:QuantitativeEta}}
\end{table*}

Nevertheless, as expected, the spectral content of the Galactic Cirrus
is not sufficiently extended towards high frequencies to provide an
excitation that is adequate for instrument identification. In
contrast, the Galactic Cirrus with point sources is more extended and
estimations are more accurate.
Tab.~\ref{tab:QuantitativeEta} presents quantitative assessments.  The
main result is that the estimation error is about $6\%$. It is a
remarkable result given the difficulty of the problem (triple problem,
complex relations, ill-possedness, and high resolution) and given that
the prior uncertainty is about $33\%$. In other words, the method is
able to capture information about instrument parameter, jointly with
noise level, regularity level, and map from a unique observation.
However, the parameter $\eta$ seems to be slightly underestimated
which, we explain as follows. The input map (with point sources)
presents a broad spectral extent whereas the \prior favours spatially
extended maps (dominated by relatively low frequencies), so the
posterior advocates a narrower PSF to compensate for this spectral
discrepancy.

\begin{figure*}[hbtp]
  \centering

  \subfigure[Proposed case~1~~~]{%
    \includegraphics[width=0.32\textwidth]{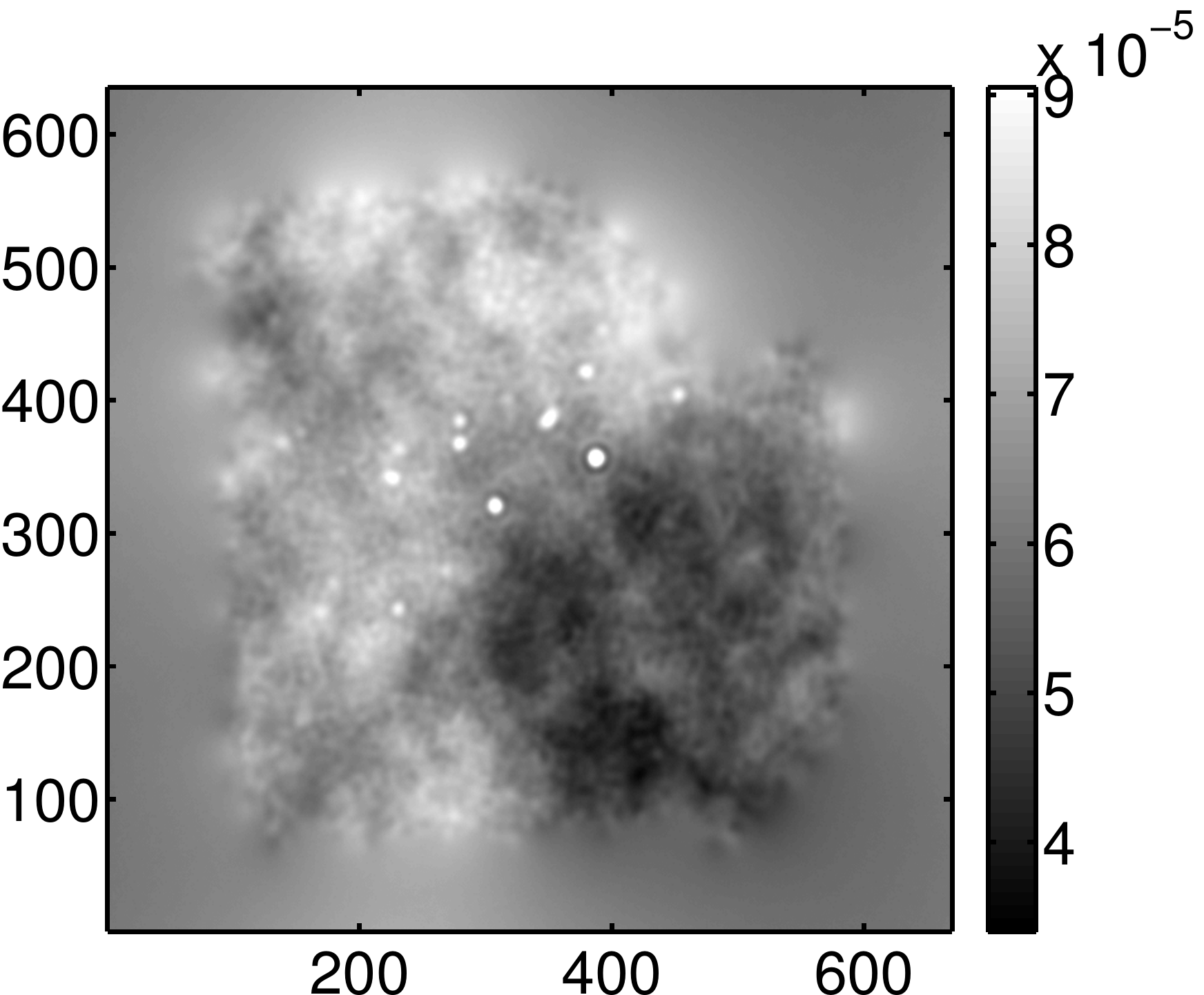}%
    \label{fig:MyopicPicsImageI}
  }
  ~~~~~
  \subfigure[Proposed case~2~~~]{%
    \includegraphics[width=0.32\textwidth]{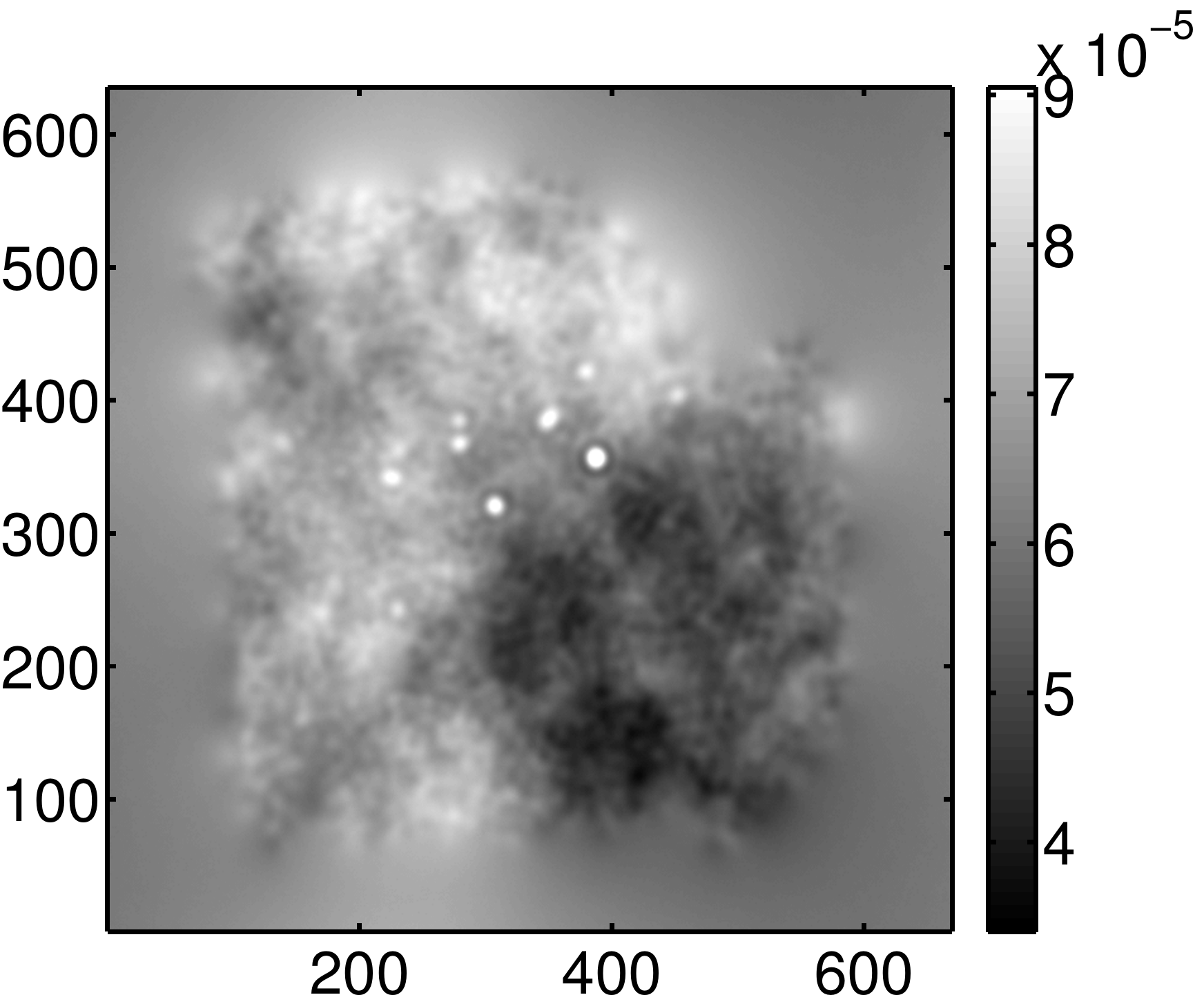}%
    \label{fig:MyopicPicsImageII}}

  \subfigure[True map]{\includegraphics[width=0.33\textwidth]%
    {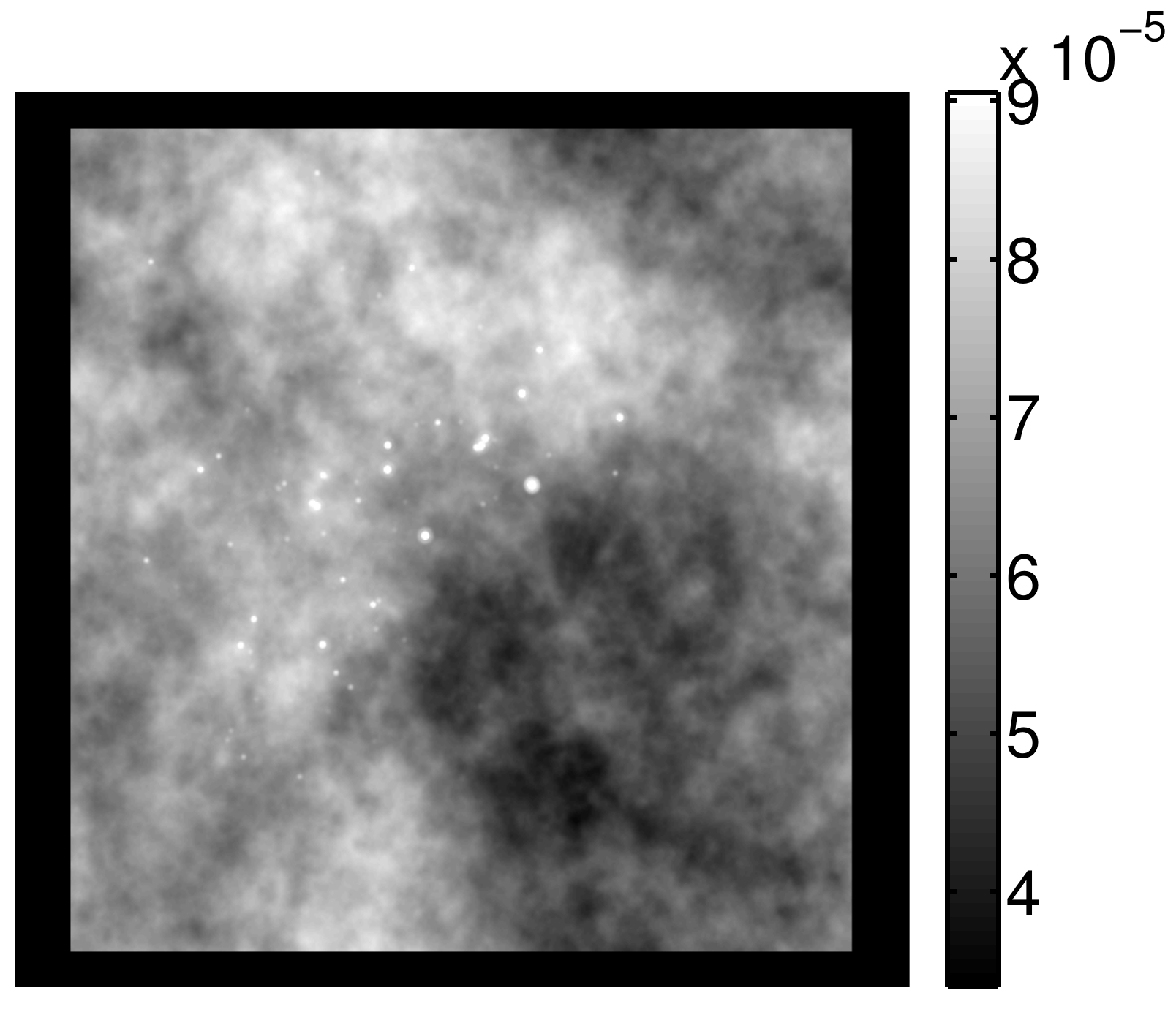}\label{fig:cirrusDotTrue}}
  \subfigure[Best]{\includegraphics[width=0.32\textwidth]%
    {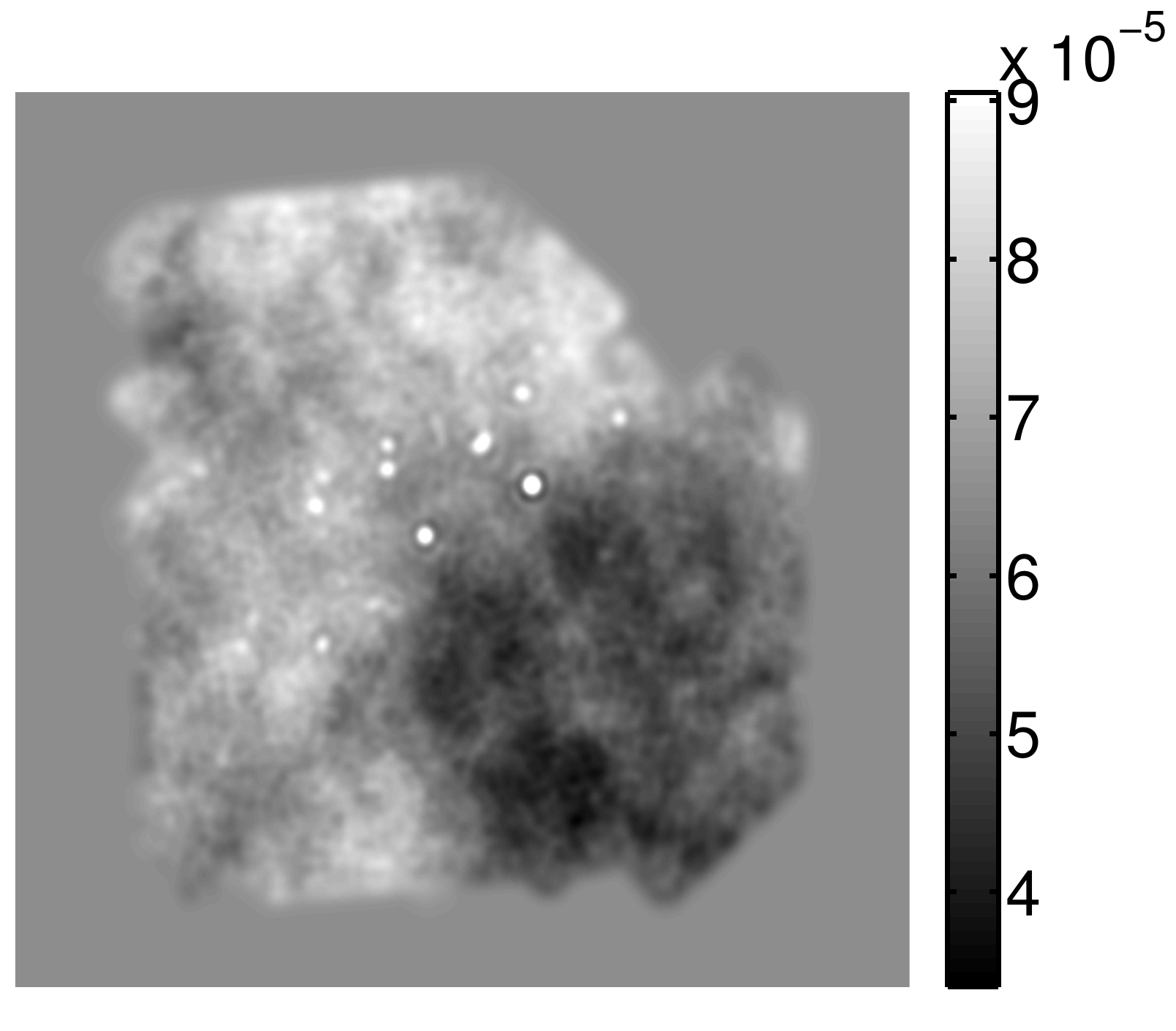}\label{fig:bestCirrusDot}}
  \subfigure[Naive map]{\includegraphics[width=0.33\textwidth]%
    {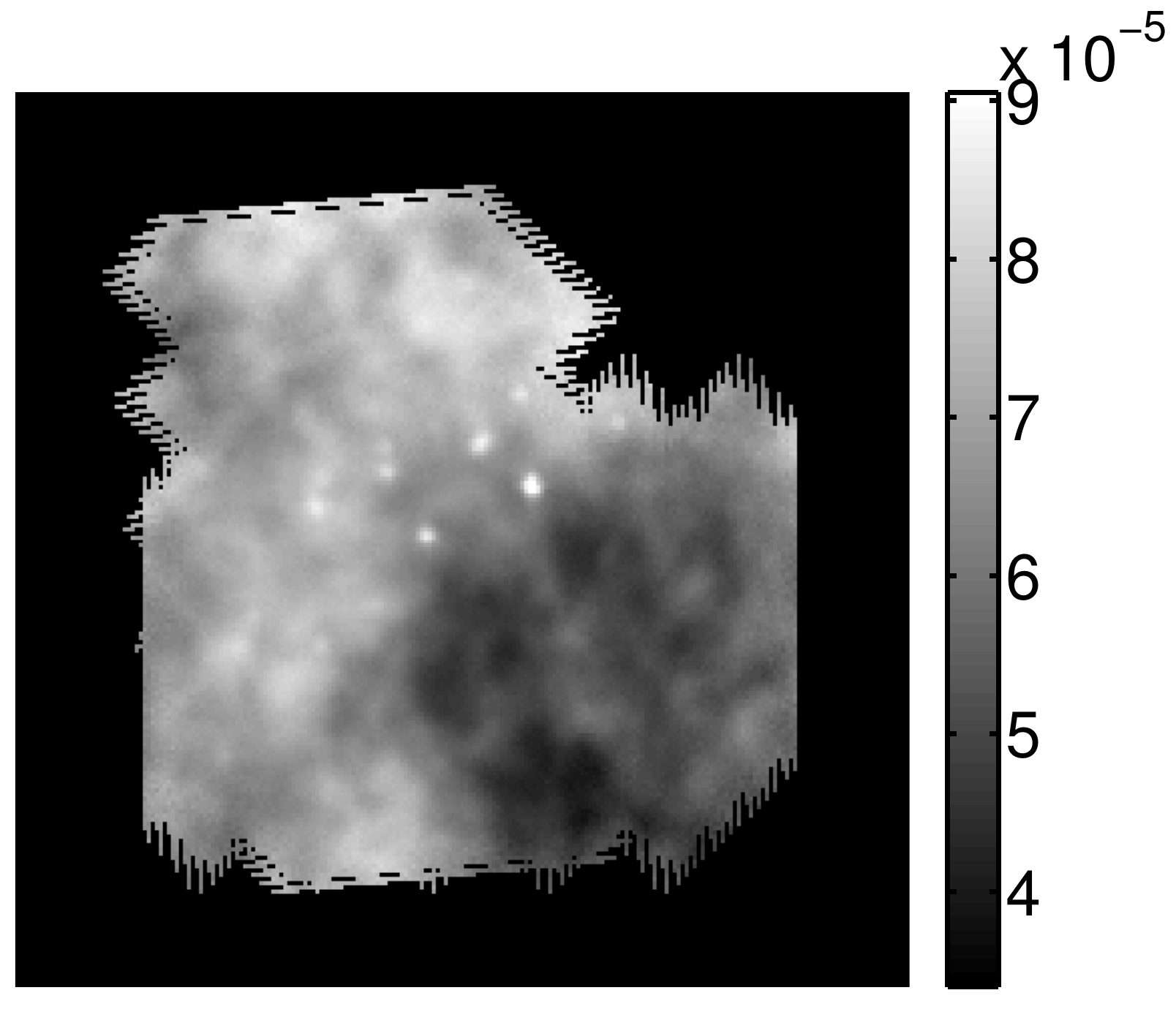}\label{fig:coaddCirrusDot}}

  \caption{Restoration of cirrus superimposed on point sources. The
    proposed maps must be compared to the maps restored with the true
    instrument parameter and the best hyperparameter and with the
    naive map. \label{fig:cirrusDot}}

\end{figure*}

Figs.~\ref{fig:MyopicPicsImageI} and~\ref{fig:MyopicPicsImageII} show
the related maps. They must be compared to the map restored with the
true instrument parameter and the best hyperparameter presented in
Fig.~7(b) of \citep{Orieux12} and in Fig.~\ref{fig:bestCirrusDot}
here. They must also be compared to the true map and the naive map
also given in \citep{Orieux12} and in
Figs.~\ref{fig:cirrusDotTrue}-\ref{fig:coaddCirrusDot} here.
As previously, the proposed maps show a fine grainy texture but
despite this defect, they remain similar to the true map. The quality
of the proposed maps is similar to the quality of the map restored
with the true instrument parameter and the best hyperparameter. In
addition, several point sources of the true map are visible on the
proposed maps but not on the naive map. In other words, the proposed
method automatically determines instrument parameter and
hyperparameters that produce a map almost as good as the best one and
better than the naive map.

\section{Conclusion}\label{Sec:Conclusion}

We described regularized methods for image reconstruction and focused
on parameter estimation:
\begin{itemize}
\item hyperparameters, which guide the trade-off between prior-based
  and observation-based information,
\item instrument parameter, which tunes the physical characteristics
  of the model of the acquisition system.
\end{itemize}
They were jointly estimated with the map of interest. We were
therefore dealing with an \emph{unsupervised and myopic inverse
  problem}.

The most delicate point is jointly handles the different types of
variables and their interactions in direct terms but, above all, in
inverse terms. From a methodology point of view, we worked in the
framework of hierarchical full Bayes strategies that model the
available information for each set of variables (map, hyperparameters,
instrument parameter, and observations) under a probability
density. We defined the \post density, which gathers the information
on the map of interest and the parameters, given the observations. We
then defined the \post mean as an estimate of the map and the \post
standard deviation as a measure of uncertainty, which gives an
uncertainty map. This approach makes it possible to work in a global
and consistent framework to solve the problem as a whole. It draws its
inspiration from our earlier works on deconvolution~\citep{Orieux10a}
and adapts them to the case at hand.

The \post density was explored by stochastic sampling using a Gibbs
algorithm. The sampling of the map was difficult: we are dealing with
a large-sized multivariate normal density for which classical
techniques do not apply. We overcame this difficulty by constructing a
sample as the minimizer of a well-chosen perturbated
criterion~\citep{Orieux12b}. Another problematic point is the
instrument parameter sampling: we are dealing with a very complex,
nonstandard density. This difficulty was overcome by means of a
Metropolis-Hastings step. The estimate of the map as well as the
parameters (\post mean) and the uncertainties (\post standard
deviation) were calculated numerically as empirical averages based on
the simulated samples.

We presented a first application of the developments (Bayesian
estimation method and stochastic sampling algorithm) in a real
context: the SPIRE instrument of the Herschel space observatory. The
study was essentially performed on simulated observations and has also
yielded some initial results on real observations. We concluded that
the approach is applicable and enables \emph{joint} estimation of the
map, the hyperparameters, and the instrument parameter from a
\emph{unique observation}. We showed, among other results, that the
quality of the proposed map is similar to that obtained when the
instrument parameter is known and the hyperparameters are fixed by
hand in a supervised way (using the sky truth). The method shows
remarkable results given the difficulty of the problem. It seems to us
that these initial results are particularly promising and worth
developing. They may open up many new perspectives for imaging in
astrophysics in a myopic and unsupervised framework.

\appendix

\section{Gamma probability density}
\label{Sec:GammaLaw}

The gamma pdf for $\gamma > 0$, with given parameters $a > 0$ and $b >
0$, is written
\begin{equation}\label{Eq:GammaPDF}
  \mathcal{G}(\gamma|a, b) = \frac{1}{b^a \Gamma(a)}
  \gamma^{a - 1}\exp \left( - \gamma / b \right).
\end{equation}
The following properties hold: mean is $\mathds{E}_{\mathcal{G}}
[\gamma] = a b$, variance is $\mathds{V}_{\mathcal{G}} [\gamma] =
ab^2$ and maximizer is $b (a - 1)$ if and only if $a > 1$.

\bibliographystyle{aa}
\bibliography{biben,revuedef,revueabr,name,biblio,BaseBiblioAutre,BaseBiblioGPI,BaseBiblioJFG}

\end{document}

%% file: abrmath.tex
\def\cro#1{\left[#1\right]}

\def\Exp#1{\exp\cro{#1}}

\newsavebox{\fminibox}
\newlength{\fminilength}

 \def\T{^\tD} \def\+{^\dagger}

\def\nequiv{\not\kern-.05em\equiv}
\def\egal{\kern-.5em=\kern-.5em}        
\def\propt{\kern-.2em\propto\kern-.2em} 
\def\wh#1{\widehat{#1}}                 
\def\wt#1{\widetilde{#1}} 

\def\intdouble{\int\kern-0.3em\int}
\def\inttriple{\int\kern-0.3em\int\kern-0.3em\int}

\def\rond#1{\overset{\kern-0.33em~_\circ}{#1}}
\def\rondit[#1]#2{\overset{\kern#1~_\circ}{#2}}


%% file: notations.tex
\def\conv{\star}
\def\eR{\mathds{R}}

\def\Ciel{\phi}

\def\Ta{T_{\alpha}}
\def\Tb{T_{\beta}}
\def\Ta{\delta_{\alpha}}
\def\Tb{\delta_{\beta}}

\def\Fe{F_{\sD}}

\def\Rc{R_{\cD}}

\def\T0{T_{0}}
\def\G0{G_{0}}

\def\Smc{\sigma_{\mD\cD}}
\def\Smc{\sigma_{\oD}} 

\def\whxij{\wh{x}_{ij}}
\def\xijstar{x_{ij}^{*}}

\def\gx{\gamma_{\xb}}

\def\gB{\gamma_{\nb}}

\def\ybt{\widetilde \yb}  
  \def\ybB{\yb_{B}}
\def\ybB1{\yb_{B+1}}

\def\my{\mb_{\yb}}

\def\Ncal{\mathcal{N}}

\def\Gcal{\mathcal{G}}

 \def\mxbt{\widetilde \mxb}

\def\mxbt{{\widetilde{\mb}_x}}


%% file: AA2.bbl
\begin{thebibliography}{57}
\expandafter\ifx\csname natexlab\endcsname\relax\def\natexlab#1{#1}\fi

\bibitem[{{Abergel} {et~al.}(2010){Abergel}, {Arab}, {Compi{\`e}gne}, {Kirk},
  {Ade}, {Anderson}, {Andr{\'e}}, {Baluteau}, {Bernard}, {Blagrave},
  {Bontemps}, {Boulanger}, {Cohen}, {Cox}, {Dartois}, {Davis}, {Emery},
  {Fulton}, {Gry}, {Habart}, {Huang}, {Joblin}, {Jones}, {Lagache}, {Lim},
  {Madden}, {Makiwa}, {Martin}, {Miville-Desch{\^e}nes}, {Molinari}, {Moseley},
  {Motte}, {Naylor}, {Okumura}, {Pinheiro Gon{\c c}alves}, {Polehampton},
  {Rodon}, {Russeil}, {Saraceno}, {Sauvage}, {Sidher}, {Spencer}, {Swinyard},
  {Ward-Thompson}, {White}, \& {Zavagno}}]{Abergel10}
{Abergel}, A., {Arab}, H., {Compi{\`e}gne}, M., {et~al.} 2010, \aap, 518, L96+

\bibitem[{Babacan {et~al.}(2010)Babacan, Molina, \& Katsaggelos}]{Babacan10}
Babacan, S., Molina, R., \& Katsaggelos, A. 2010, \uppercase{ieee} {T}rans.
  {I}mage {P}rocessing, 19, 53

\bibitem[{Bishop {et~al.}(2008)Bishop, Molina, \& Hopgood}]{Bishop08}
Bishop, T., Molina, R., \& Hopgood, J. 2008, in {P}roc. \uppercase{ieee}
  \uppercase{icip}

\bibitem[{Blanc {et~al.}(2003)Blanc, Mugnier, \& Idier}]{Blanc03}
Blanc, A., Mugnier, L., \& Idier, J. 2003, {J}. {O}pt. {S}oc. {A}mer. ({A}),
  20, 1035

\bibitem[{Chantas {et~al.}(2007)Chantas, Galatsanos, \& Woods}]{Chantas07}
Chantas, G.~K., Galatsanos, N.~P., \& Woods, N.~A. 2007, \uppercase{ieee}
  {T}rans. {I}mage {P}rocessing, 16, 1821

\bibitem[{Chellappa \& Chatterjee(1985)}]{Chellappa85}
Chellappa, R. \& Chatterjee, S. 1985, \uppercase{ieee} {T}rans. {A}coust.
  {S}peech, {S}ignal {P}rocessing, {ASSP}-33, 959

\bibitem[{Chellappa \& Jain(1992)}]{Chellappa92}
Chellappa, R. \& Jain, A. 1992, Markov Random Fields: Theory and Application
  (Academic Press Inc)

\bibitem[{Conan {et~al.}(1998)Conan, Mugnier, Fusco, Michau, \& G.}]{Conan98a}
Conan, J.-M., Mugnier, L., Fusco, T., Michau, V., \& G., R. 1998, Applied
  Optics, 37, 4614

\bibitem[{de~Figueiredo \& Leitao(1997)}]{Figueiredo97}
de~Figueiredo, M.~T. \& Leitao, J. M.~N. 1997, \uppercase{ieee} {T}rans.
  {I}mage {P}rocessing, 6, 1089

\bibitem[{Descombes {et~al.}(1999)Descombes, Morris, Zerubia, \&
  Berthod}]{Descombes99a}
Descombes, X., Morris, R., Zerubia, J., \& Berthod, M. 1999, \uppercase{ieee}
  {T}rans. {I}mage {P}rocessing, 8, 954

\bibitem[{Dobigeon {et~al.}(2009)Dobigeon, Hero, \& Tourneret}]{Dobigeon09}
Dobigeon, N., Hero, A., \& Tourneret, J.-Y. 2009, \uppercase{ieee} {T}rans.
  {I}mage {P}rocessing, 18

\bibitem[{F{\'e}ron(2006)}]{Feron06}
F{\'e}ron, O. 2006, Ph{D} {T}hesis, Universit\'e de Paris-Sud, Orsay, , France

\bibitem[{Fortier {et~al.}(1993)Fortier, Demoment, \& Goussard}]{Fortier93}
Fortier, N., Demoment, G., \& Goussard, Y. 1993, {J}. {V}isual {C}omm. {I}mage
  {R}epres., 4, 157

\bibitem[{Fusco {et~al.}(1999)Fusco, Véran, Conan, \& Mugnier}]{Fusco99}
Fusco, T., Véran, J.-P., Conan, J.-M., \& Mugnier, L.~M. 1999, Astron.
  Astrophys. Suppl. Ser., 134, 193­200

\bibitem[{Geman \& Yang(1995)}]{Geman95}
Geman, D. \& Yang, C. 1995, \uppercase{ieee} {T}rans. {I}mage {P}rocessing, 4,
  932

\bibitem[{Gilks {et~al.}(1996)Gilks, Richardson, \& Spiegelhalter}]{Gilks96}
Gilks, W.~R., Richardson, S., \& Spiegelhalter, D.~J. 1996, {M}arkov {C}hain
  {M}onte {C}arlo in practice (Boca Raton,: Chapman \& Hall/CRC)

\bibitem[{Giovannelli(2008)}]{Giovannelli08}
Giovannelli, J.-F. 2008, \uppercase{ieee} {T}rans. {I}mage {P}rocessing, 17, 16

\bibitem[{Golub {et~al.}(1979)Golub, Heath, \& Wahba}]{Golub79}
Golub, G.~H., Heath, M., \& Wahba, G. 1979, {T}echnometrics, 21, 215

\bibitem[{{Griffin} {et~al.}(2010){Griffin}, {Abergel}, {Abreu}, {Ade},
  {Andr{\'e}}, {Augueres}, {Babbedge}, {Bae}, {Baillie}, {Baluteau}, {Barlow},
  {Bendo}, {Benielli}, {Bock}, {Bonhomme}, {Brisbin}, {Brockley-Blatt},
  {Caldwell}, {Cara}, {Castro-Rodriguez}, {Cerulli}, {Chanial}, {Chen},
  {Clark}, {Clements}, {Clerc}, {Coker}, {Communal}, {Conversi}, {Cox},
  {Crumb}, {Cunningham}, {Daly}, {Davis}, {de Antoni}, {Delderfield}, {Devin},
  {di Giorgio}, {Didschuns}, {Dohlen}, {Donati}, {Dowell}, {Dowell}, {Duband},
  {Dumaye}, {Emery}, {Ferlet}, {Ferrand}, {Fontignie}, {Fox}, {Franceschini},
  {Frerking}, {Fulton}, {Garcia}, {Gastaud}, {Gear}, {Glenn}, {Goizel},
  {Griffin}, {Grundy}, {Guest}, {Guillemet}, {Hargrave}, {Harwit}, {Hastings},
  {Hatziminaoglou}, {Herman}, {Hinde}, {Hristov}, {Huang}, {Imhof}, {Isaak},
  {Israelsson}, {Ivison}, {Jennings}, {Kiernan}, {King}, {Lange}, {Latter},
  {Laurent}, {Laurent}, {Leeks}, {Lellouch}, {Levenson}, {Li}, {Li},
  {Lilienthal}, {Lim}, {Liu}, {Lu}, {Madden}, {Mainetti}, {Marliani}, {McKay},
  {Mercier}, {Molinari}, {Morris}, {Moseley}, {Mulder}, {Mur}, {Naylor},
  {Nguyen}, {O'Halloran}, {Oliver}, {Olofsson}, {Olofsson}, {Orfei}, {Page},
  {Pain}, {Panuzzo}, {Papageorgiou}, {Parks}, {Parr-Burman}, {Pearce},
  {Pearson}, {P{\'e}rez-Fournon}, {Pinsard}, {Pisano}, {Podosek}, {Pohlen},
  {Polehampton}, {Pouliquen}, {Rigopoulou}, {Rizzo}, {Roseboom}, {Roussel},
  {Rowan-Robinson}, {Rownd}, {Saraceno}, {Sauvage}, {Savage}, {Savini},
  {Sawyer}, {Scharmberg}, {Schmitt}, {Schneider}, {Schulz}, {Schwartz},
  {Shafer}, {Shupe}, {Sibthorpe}, {Sidher}, {Smith}, {Smith}, {Smith},
  {Spencer}, {Stobie}, {Sudiwala}, {Sukhatme}, {Surace}, {Stevens}, {Swinyard},
  {Trichas}, {Tourette}, {Triou}, {Tseng}, {Tucker}, {Turner}, {Vaccari},
  {Valtchanov}, {Vigroux}, {Virique}, {Voellmer}, {Walker}, {Ward}, {Waskett},
  {Weilert}, {Wesson}, {White}, {Whitehouse}, {Wilson}, {Winter}, {Woodcraft},
  {Wright}, {Xu}, {Zavagno}, {Zemcov}, {Zhang}, \& {Zonca}}]{Griffin10}
{Griffin}, M.~J., {Abergel}, A., {Abreu}, A., {et~al.} 2010, \aap, 518, L3+

\bibitem[{Hansen(1992)}]{Hansen92}
Hansen, P. 1992, \uppercase{siam} {R}ev., 34, 561

\bibitem[{Idier(2008)}]{Idier08}
Idier, J., ed. 2008, Bayesian Approach to Inverse Problems (London: ISTE Ltd
  and John Wiley \& Sons Inc.)

\bibitem[{Jalobeanu {et~al.}(2002)Jalobeanu, Blanc-Feraud, \&
  Zerubia}]{Jalobeanu02a}
Jalobeanu, A., Blanc-Feraud, L., \& Zerubia, J. 2002, in {P}roc.
  \uppercase{ieee} \uppercase{icassp}, Vol.~4, 3580--3583

\bibitem[{Kass \& Wasserman(1996)}]{Kass96}
Kass, R.~E. \& Wasserman, L. 1996, {J}. {A}mer. {S}tatist. {A}ssoc., 91, 1343

\bibitem[{Lalanne {et~al.}(2001)Lalanne, Pr{\'e}vost, \& Chavel}]{Lalanne01}
Lalanne, P., Pr{\'e}vost, D., \& Chavel, P. 2001, {A}pplied {O}ptics, 40

\bibitem[{Lam \& Goodman(2000)}]{Lam00}
Lam, E.~Y. \& Goodman, J.~W. 2000, J. Opt. Soc. Am. A, 17, 1177

\bibitem[{Lanterman {et~al.}(2000)Lanterman, Grenander, \&
  Miller}]{Lanterman00}
Lanterman, A.~D., Grenander, U., \& Miller, M.~I. 2000, \uppercase{ieee}
  {T}rans. {P}attern {A}nal. {M}ach. {I}ntell., 22, 337

\bibitem[{Li(2001)}]{Li01}
Li, S.~Z. 2001, Markov Random Field Modeling in Image Analysis (Tokyo:
  Springer-Verlag)

\bibitem[{Likas \& Galatsanos(2004)}]{Likas04}
Likas, A.~C. \& Galatsanos, N.~P. 2004, \uppercase{ieee} {T}rans. {I}mage
  {P}rocessing, 52, 2222

\bibitem[{Mallat(2008)}]{Mallat08}
Mallat, S. 2008, A Wavelet Tour of Signal Processing: The Sparse Way (Academic
  Press Inc.)

\bibitem[{Molina {et~al.}(1999)Molina, Katsaggelos, \& Mateos}]{Molina99}
Molina, R., Katsaggelos, A.~K., \& Mateos, J. 1999, \uppercase{ieee} {T}rans.
  {I}mage {P}rocessing, 8, 231

\bibitem[{Molina {et~al.}(2006)Molina, Mateos, \& Katsaggelos}]{Molina06}
Molina, R., Mateos, J., \& Katsaggelos, A.~K. 2006, \uppercase{ieee} {T}rans.
  {I}mage {P}rocessing, 15, 3715

\bibitem[{Mugnier {et~al.}(2004)Mugnier, Fusco, \& Conan}]{Mugnier04}
Mugnier, L., Fusco, T., \& Conan, J.-M. 2004, {J}. {O}pt. {S}oc. {A}mer., 21,
  1841

\bibitem[{Ocvirk {et~al.}(2006)Ocvirk, Pichon, Lan\c{c}on, \&
  Thi\'ebaut}]{Pichon06}
Ocvirk, P., Pichon, C., Lan\c{c}on, A., \& Thi\'ebaut, E. 2006, {M}onthly
  {N}otices of the {R}oyal {A}stronomical {S}ociety, 365, 46

\bibitem[{Orieux {et~al.}(2012{\natexlab{a}})Orieux, F{\'e}ron, \&
  Giovannelli}]{Orieux12b}
Orieux, F., F{\'e}ron, O., \& Giovannelli, J.-F. 2012{\natexlab{a}},
  \uppercase{ieee} {S}ignal {P}rocessing {L}etters

\bibitem[{Orieux {et~al.}(2010{\natexlab{a}})Orieux, Giovannelli, \&
  Rodet}]{Orieux10}
Orieux, F., Giovannelli, J.-F., \& Rodet, T. 2010{\natexlab{a}}, {J}. {O}pt.
  {S}oc. {A}mer., 27, 1593

\bibitem[{Orieux {et~al.}(2010{\natexlab{b}})Orieux, Giovannelli, \&
  Rodet}]{Orieux10a}
Orieux, F., Giovannelli, J.-F., \& Rodet, T. 2010{\natexlab{b}}, in {P}roc.
  \uppercase{ieee} \uppercase{icassp}, Dallas,

\bibitem[{Orieux {et~al.}(2012{\natexlab{b}})Orieux, Giovannelli, Rodet,
  Ayasso, \& Abergel}]{Orieux12}
Orieux, F., Giovannelli, J.-F., Rodet, T., Ayasso, H., \& Abergel, A.
  2012{\natexlab{b}}, {A}stron. {A}strophys.

\bibitem[{Orieux {et~al.}(2009)Orieux, Rodet, \& Giovannelli}]{Orieux09a}
Orieux, F., Rodet, T., \& Giovannelli, J.-F. 2009, in {P}roc. \uppercase{ieee}
  \uppercase{icip}, Le Caire, Egypt

\bibitem[{Pankajakshani {et~al.}(2009)Pankajakshani, Zhang, Blanc-F\'{e}raud,
  Kam, Olivo-Marin, \& Zerubia}]{Pankajakshan09}
Pankajakshani, P., Zhang, B., Blanc-F\'{e}raud, L., {et~al.} 2009, {A}pplied
  {O}ptics, 48, 4437

\bibitem[{Pascazio \& Ferraiuolo(2003)}]{Pascazio03}
Pascazio, V. \& Ferraiuolo, G. 2003, \uppercase{ieee} {T}rans. {I}mage
  {P}rocessing, 12, 572

\bibitem[{{Pilbratt} {et~al.}(2010){Pilbratt}, {Riedinger}, {Passvogel},
  {Crone}, {Doyle}, {Gageur}, {Heras}, {Jewell}, {Metcalfe}, {Ott}, \&
  {Schmidt}}]{Pilbratt10}
{Pilbratt}, G.~L., {Riedinger}, J.~R., {Passvogel}, T., {et~al.} 2010, \aap,
  518, L1+

\bibitem[{Robert(2005)}]{Robert05}
Robert, C.~P. 2005, The {B}ayesian {C}hoice, Statistiques et probabilit\'es
  appliqu\'ees (Paris, France: Springer)

\bibitem[{Robert \& Casella(2004)}]{Robert04}
Robert, C.~P. \& Casella, G. 2004, {M}onte-{C}arlo Statistical Methods,
  Springer Texts in Statistics (New York, \sca{ny},: Springer)

\bibitem[{Rodet {et~al.}(2008)Rodet, Orieux, Giovannelli, \& Abergel}]{Rodet08}
Rodet, T., Orieux, F., Giovannelli, J.-F., \& Abergel, A. 2008,
  \uppercase{ieee} {J}. of {S}elec. {T}opics in {S}ignal {P}roc., 2, 802

\bibitem[{Rue(2001)}]{Rue01}
Rue, H. 2001, {J}. {R}. {S}tatist. {S}oc. {B}, 63

\bibitem[{Rue \& Held(2005)}]{Rue05}
Rue, H. \& Held, L. 2005, Monographs on Statistics and Applied Probability,
  Vol. 104, Gaussian {M}arkov Random Fields: {T}heory and Applications (Chapman
  \& Hall)

\bibitem[{Saquib {et~al.}(1998)Saquib, Bouman, \& Sauer}]{Saquib98}
Saquib, S.~S., Bouman, C.~A., \& Sauer, K.~D. 1998, \uppercase{ieee} {T}rans.
  {I}mage {P}rocessing, 7, 1029

\bibitem[{Tan {et~al.}(2010)Tan, Li, \& Stoica}]{Tan10}
Tan, X., Li, J., \& Stoica, P. 2010, in {P}roc. \uppercase{ieee}
  \uppercase{icassp}, 3634 --3637

\bibitem[{Thi\'ebaut(2008)}]{Thiebaut08}
Thi\'ebaut, E. 2008, in proc. {SPIE}: Astronomical Telescopes and
  Instrumentation, Vol. 7013, 70131--I

\bibitem[{Thiébaut \& Conan(1995)}]{Thiebaut95}
Thiébaut, E. \& Conan, J.-M. 1995, {J}. {O}pt. {S}oc. {A}mer. ({A}), 12, 485

\bibitem[{Tikhonov \& Arsenin(1977)}]{Tikhonov77}
Tikhonov, A. \& Arsenin, V. 1977, Solutions of Ill-Posed Problems (Washington,
  \sca{dc}: Winston)

\bibitem[{Vacar {et~al.}(2011)Vacar, Giovannelli, \& Berthoumieu}]{Vacar11}
Vacar, C., Giovannelli, J.-F., \& Berthoumieu, Y. 2011, in {P}roc.
  \uppercase{ieee} \uppercase{icassp}, Prague, Czech Republic

\bibitem[{{Wiegelmann} \& {Inhester}(2003)}]{Wiegelmann03}
{Wiegelmann}, T. \& {Inhester}, B. 2003, {S}olar {P}hysics, 214, 287

\bibitem[{Winkler(2003)}]{Winkler03}
Winkler, G. 2003, Image Analysis, Random Fields and Markov Chain Monte Carlo
  Methods (Springer Verlag, Berlin Germany)

\bibitem[{Xu \& Lam(2009)}]{Xu09}
Xu, Z. \& Lam, E.~Y. 2009, Opt. Lett., 34, 1453

\bibitem[{Zhang {et~al.}(2007)Zhang, Zerubia, \& Olivo-Marin}]{Zhang06}
Zhang, B., Zerubia, J., \& Olivo-Marin, J.-C. 2007, {A}pplied {O}ptics, 46,
  1819

\bibitem[{Zhou {et~al.}(1997)Zhou, Leahy, \& Qi}]{Zhou97}
Zhou, Z., Leahy, R.~M., \& Qi, J. 1997, \uppercase{ieee} {T}rans. {I}mage
  {P}rocessing, 6, 844

\end{thebibliography}
